\documentclass{article}
\usepackage{jheppub}
 \normalsize

\def\lsim{\raise0.3ex\hbox{$\;<$\kern-0.75em\raise-1.1ex\hbox{$\sim\;$}}}
\def\gsim{\raise0.3ex\hbox{$\;>$\kern-0.75em\raise-1.1ex\hbox{$\sim\;$}}}

\usepackage{graphics}
\usepackage{rotating}
\usepackage{multirow}
\usepackage{pstricks}
\usepackage{dcolumn}
\usepackage{bbold}
\usepackage{natbib}
\def\be{\begin{equation}}
\def\ee{\end{equation}}
\def\bea{\begin{eqnarray}}
\def\eea{\end{eqnarray}}

\usepackage{epsfig}
\usepackage{slashed}
 \normalsize

\begin{document}
\title{Large $BR(h \to \tau \mu)$ in Supersymmetric Models }
\author{ A. Hammad$^1$, S. Khalil$^1$ and C. Un$^{1,2}$ }
\vspace*{0.2cm}
\affiliation{$^1$Center for Fundamental Physics, Zewail City of Science and Technology, 6 October City, Giza, Egypt. \\
$^2$Department of Physics, Uluda\~{g} University, 16059, T\"{u}rkiye}
\date{\today}

\abstract{

We analyze the Lepton Flavor Violating (LFV) Higgs decay $h\to \tau \mu$ in three supersymmetric models: Minimal Supersymmetric Standard Model (MSSM), Supersymmetric Seesaw Model (SSM), and Supersymmetric $B-L$ model with Inverse Seesaw (BLSSM-IS).  We show that in generic MSSM, with non-universal slepton masses and/or trilinear couplings, it is not possible to enhance $BR(h \to \tau \mu)$ without violating the experimental bound on the $BR(\tau \to \mu \gamma)$. In SSM, where flavor mixing is radiatively generated, the LFV process $\mu \to e \gamma$ strictly constrains the parameter space and the maximum value of $BR(h \to \tau \mu)$ is of order $10^{-10}$, which is extremely smaller than the recent results reported by the CMS and ATLAS experiments. In BLSSM-IS, with universal soft SUSY breaking terms at the grand unified scale, we emphasize  that the measured values of $BR(h \to \tau \mu)$ can be accommodated in a wide region of parameter space without violating LFV constraints. 
Thus, confirming the LFV Higgs decay results will be a clear signal of BLSSM-IS type of models.  Finally, the signal of $h \to \tau \mu$ in the BLSSM-IS at the LHC, which has a tiny background, is analyzed. 

}

\maketitle

\section{Introduction}
\label{sec:intro}

The CMS  and ATLAS  collaborations reported the first signal of LFV Higgs decay $h \to \tau \mu$.  The branching ratio of this decay is found as \cite{Khachatryan:2015kon,Aad:2015gha}
\bea
BR(h\to \tau \mu) &=& \left( 8.4^{+3.9}_{-3.7} \right) \times 10^{-3} ~~~~~~ {\rm (CMS)},\\
BR(h\to \tau \mu) &=& \left( 7.7 \pm 6.2 \right) \times 10^{-3} ~~~~~~ {\rm (ATLAS)}.
\eea
The Standard Model (SM) predicts that there should be no tree-level LFV Higgs coupling at the renormalizable level. These LFV processes are forbidden also because of the lepton flavor symmetry which emerges accidentally in the SM. On the other hand, many extensions of the SM do not exhibit such symmetries, and therefore, the measurements on the LFV processes can provide an indirect signal for the new physics beyond the SM, in particular SUSY models. In the MSSM framework, even if one assumes no mixing in the lepton sector, a misalignment in the slepton sector with the soft SUSY breaking (SSB) terms can induce LFV processes through the loop processes mediated by charginos or neutralinos. Besides an ad-hoc assumption of a possible misalignment in the MSSM sector, one can also realize non-zero mixing among the slepton families by extending the MSSM with right-handed neutrinos. In the presence of right-handed neutrinos, LFV processes are also favored by the neutrino oscillations \cite{Arganda:2004bz}.

However, SUSY models with non-zero family mixing in the sleptons also result in enhancement in other LFV processes such as $\mu \rightarrow e \gamma$, $\tau\rightarrow e \gamma$, and $\tau \rightarrow \mu \gamma$. The experimental exclusion limits on these processes are established as $BR(\mu\rightarrow e \gamma)\leq 5.7\times 10^{-13}$ \cite{Adam:2013mnn}, $BR(\tau\rightarrow e \gamma)\leq 3.3\times 10^{-8}$, and $BR(\tau\rightarrow \mu \gamma)\leq 4.4\times 10^{-8}$ \cite{Aubert:2009ag}. These measurements, especially those on $\mu\rightarrow e \gamma$, provide severe constraints on these models, that may lead to a sizable enhancement in LFV higgs decays. 

In this work, we consider the LFV Higgs decays in several SUSY models in light of the current experimental constraints on the LFV processes mentioned above. We first consider a generic MSSM framework in which the slepton mass matrix is, in general, non-diagonal. If one assumes only the relevant terms contributing to $h\rightarrow \tau \mu$ to be non-zero, and sets the others zero for simplicity, the bounds from the LFV processes $\mu \rightarrow e \gamma$ and $\tau \rightarrow e \gamma$ can easily be escaped. However,  the terms that enhance $BR(h\rightarrow \tau \mu)$ also contribute to $\tau \rightarrow \mu \gamma$, and it is not possible to realize testable results in respect of measurements of the LFV Higgs boson decays without violating the experimental bound on the $BR(\tau \rightarrow \mu \gamma)$. It is established that the family mixing in the lepton sector can be based on seesaw mechanisms in the presence of right-handed neutrinos. In this case, even though it is possible to enhance the LFV Higgs boson decays, it also opens all the other mentioned LFV processes and $\mu \rightarrow e \gamma$ strictly constrains the parameter space \cite{Calibbi:2012gr,Masiero:2004hg,Masiero:2002jn}. In this respect, the SUSY models with seesaw mechanisms are worth to explore.  

The rest of the paper is organized as follows: In Section \ref{sec:MSSM}, we first consider the status of LFV processes in a generic MSSM framework. Then, we investigate the implications of SUSY models with the right-handed neutrinos in Section \ref{sec:SS1}, in which the neutrinos acquire a non-zero masses via SSM mechanism. Finally we explore BLSSM-IS in Section \ref{sec:BLSSMISS}. We impose the universal boundary conditions at the grand unified scale ($M_{{\rm GUT}}$), and analyze the low scale implications calculated through the renormalization group equations (RGEs) run down to the electroweak scale. Section \ref{sec:signal} represents detection of possible signals over the relevant background at the LHC by considering two benchmark points. At the end, we summarize our results and conclude in Section \ref{sec:conc}.

\section{$h \to \tau \mu$ in the MSSM}
\label{sec:MSSM}
As in the SM, the neutrinos are massless in the MSSM, hence a diagonal  lepton Yukawa matrix can be adopted.  In this regard,  possible LFV sources may be generated from the SSB sector. In particular, the LFV in the MSSM stem from the off-diagonal elements of slepton mass matrix, which can be parametrized as \cite{Paradisi:2006jp}
\bea {\cal M}^2_{\tilde{L}} &= & \left(
\begin{array}{cc} V_L^\ell m^2_{\tilde{\ell}_L} V_L^{\ell\dagger} + m_\ell^2  + \frac{\cos
2\beta}{2}(M_Z^2 - 2M_W^2)  ~~~~& ~~~~ v \cos \beta V_L^\ell T_\ell^{*} V_R^{\ell\dagger} - \tan\beta \mu m_\ell\\
v \cos \beta V_R^\ell T_\ell^T V_L^{\ell\dagger} - \tan\beta \mu^{*} m_\ell  ~~~~& ~~~~ V_R^\ell m^2_{\tilde{\ell}^c_L}
V_R^{\ell\dagger} + m_\ell^2  - \cos
2\beta M_Z^2\sin^2\theta_W
\end{array}\right)\!, ~~~~
\label{mLL2}
\eea
where $V_L^\ell$ and $V_R^\ell$ are unitary matrices that diagonal the lepton mass matrix $m_\ell$. $m^2_{\tilde{\ell}_L} $ and $m^2_{\tilde{\ell}^c_L}$ are the mass matrices for the left-handed sleptons and right-handed sleptons. The trilinear coupling, $T_\ell$, is usually given in terms of the lepton Yukawa coupling times a dimension full $A$-terms, {\it i.e.},  $(T^\ell)_{ij} = (Y^\ell)_{ij} (A^\ell)_{ij}$. Since the MSSM has no right handed neutrino superfield, the sneutrino mass matrix takes the form
\bea
{\cal M}^2_{\tilde{\nu}_L}&=& V^\ell_L m^2_{\tilde{\ell}_L}
V_L^{\ell\dagger} \!+\! \frac{\cos 2\beta}{2} M_Z^2,
\label{eq:sfmass} 
\eea
The two matrices ${\cal M}^2_{\tilde{L}}$ and ${\cal
M}^2_{\tilde{\nu}_L}$  can be diagonalised by two unitary matrices
$Z^{\tilde{\nu}}$ and $Z^{\tilde{\ell}}$
  \bea
M^2_{\tilde{\nu}}&=&
Z^{\tilde{\nu}} {\cal M}^2_{\tilde{\nu}_L} {Z^{\tilde{\nu}}}^\dagger,\nonumber\\
M^2_{\tilde{\ell}} &=& Z^{\tilde{\ell}} {\cal M}^2_{\tilde{L}}{Z^{\tilde{\ell}}}^\dagger.
 \eea

\begin{figure}[ht!]
\centering
\includegraphics[scale = 0.4]{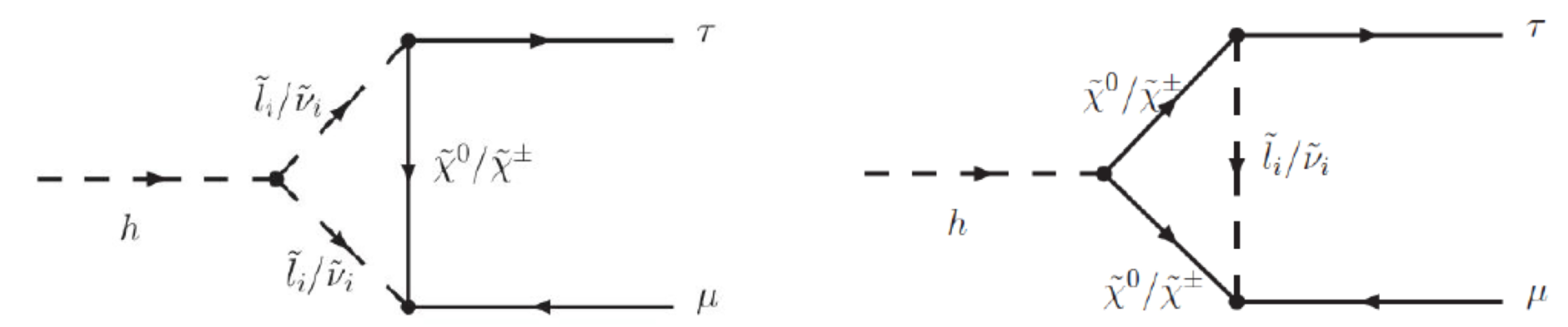}
\caption{$h \to \tau \mu$ in the MSSM through chargino and sneutrino or neutralino and charged slepton exchanges}
\label{feynman}
\end{figure}

In this context, the LFV decay $h\rightarrow \tau \mu$ can occur at one loop level via mediation of neutralinos or charginos \cite{Aloni:2015wvn} as shown in Figure \ref{feynman}. The Lagrangian for the interactions among the leptons, sleptons and neutralino or chargino is given as 

\begin{eqnarray}
\mathcal{L}_{\rm int} = \sum_{i=1}^{2}\overline{\ell}_{i}\left( \Gamma^{\chi_j^- \ell_i \tilde{\nu}_k}_L P_{L}+\Gamma^{\chi_j^- \ell_i \tilde{\nu}_k}_R P_{R}\right) \tilde{\chi }_{j}^{-}\tilde{\nu}_{k} +\sum_{j=1}^{4}\overline{\ell}%
_{i}\left(\Gamma_L^{\chi^0_j \ell_i \tilde{l}_k} P_{L}+ \Gamma_R^{\chi^0_j \ell_i \tilde{l}_k} P_{R}\right)\tilde{\chi }^0_{j} \tilde{\ell}_{k}+h.c. ,  ~~~~
\label{gauginoints}
\end{eqnarray}
where
\bea
\Gamma^{\chi^-_j \ell_i \tilde{\nu}_k}_L &=& i U^\ast_{j2} \sum^{3}_{b=1} Z^{\tilde{\nu}^\ast}_{kb} Y^\ell_{ib},\\
\Gamma^{\chi^+_j \ell_i \tilde{\nu}_k}_R &=& -i g_2V_{j1} Z^{\tilde{\nu}}_{ki},\\
\Gamma^{\chi^0_j \ell_i \tilde{\ell}_k}_L &=& \frac{i}{2}\left[ 2N_{j3} \sum^3_{a=1} Y^\ell_{ai}Z^{\tilde{\ell}^\ast}_{k,a+3}- \sqrt{2} \left(g_1 N_{j1} + g_2 N_{j2}\right)  Z^{\tilde{\ell}^\ast}_{ki} \right], ~~~\\
\Gamma^{\chi_k^0 \ell_i \tilde{\ell}_k}_R &=& i \left[ \sqrt{2}g_1 Z^{\tilde{\ell}^\ast}_{k,3+i} N^\ast_{j1} + \sum^3_{b=1}Y^{\ell^\ast}_{ib} Z^{\tilde{\ell}^\ast}_{kb}N_{j3}\right].
\label{gaugevertices}
\eea

Furthermore, the interactions of the SM like Higgs boson with sleptons, sneutrinos, neutralinos and charginos complete the analysis over the loop integrals given in Figure \ref{feynman}. They are given by \cite{Rosiek:1995kg}

\bea
\Gamma^{h \tilde{\nu} \tilde{\nu}^\ast} &=& -\frac{i}{4} ( g^2_1+g^2_2) \left[\upsilon_d Z^H_{11} - \upsilon_u Z^H_{12} \right],\\
\Gamma^{h \tilde{\ell}_j \tilde{\ell}_k^\ast} &=& \frac{-i}{4}(g^2_1-g^2_2)\sum^3_{a=1}Z^{\tilde{\ell}^\ast}_{ja}Z^{\tilde{\ell}}_{ka} \left(\upsilon_d Z^H_{11} - \upsilon_u Z^H_{12} \right) - \frac{g^2_1}{2} \sum^3_{a=1}Z^{\tilde{\ell}^\ast}_{j,3+a}Z^{\tilde{\ell}}_{k,3+a}\left(\upsilon_uZ^H_{12} -\upsilon_dZ^H_{11} \right) 
\nonumber\\
&-&\frac{i}{\sqrt{2}}\sum^3_{a,b=1}\left( Z^{\tilde{\ell}^\ast}_{jb} Z^{\tilde{\ell}}_{k,3+a} + Z^{\tilde{\ell}^\ast}_{j,3+a}Z^{\tilde{\ell}}_{ka} \right) T^\ell_{ab}Z^H_{11} 
+ \frac{i}{\sqrt{2}}\mu \sum^3_{a,b=1} \left(Z^{\tilde{\ell}^\ast}_{jb}  Z^{\tilde{\ell}}_{k,3+a} + Z^{\tilde{\ell}^\ast}_{j,3+a}Z^{\tilde{\ell}}_{kb}\right) Y^{\ell}_{ab} Z^H_{12}, \nonumber\\ \\
\Gamma^{h \tilde{\chi}^+_i \tilde{\chi}^-_j}_L &=& -i\frac{g_2}{\sqrt{2}}\left(U^\ast_{i1}V^\ast_{j1}Z^H_{12}+U^\ast_{j1}V^\ast_{i1}Z^H_{11} \right),\\
\Gamma^{h \tilde{\chi}^0_i \tilde{\chi}^0_j}_L &=& \frac{i}{2}\Big[N^\ast_{i1}(g_1N^\ast_{j1}-g_2N^\ast_{j2})Z^H_{11}-g_2N^\ast_{i2}N^\ast_{j3}Z^H_{11}-g_1N^\ast_{i4}N^\ast_{j2}Z^H_{12}-g_1N^\ast_{i4}N^\ast_{j2}Z^h_{12}+g_2N^\ast_{i4}N^\ast_{j2}Z^H_{12}\nonumber\\
&+& g_2N^\ast_{i2}N^\ast_{j4}Z^H_{12}+g_1N^\ast_{i2}(N^\ast_{j3}Z^H_{11}-N^\ast_{j4}Z^H_{12})\Big].
\eea
From these expressions one finds that the vertex $\Gamma^{h \tilde{\ell}_2 \tilde{\ell}_3^\ast}$ can be enhanced through the last two terms, which are proportional to $T^\ell_{ab}$ and $\mu Y^\ell_{ab}$, respectively. The trilinear coupling $T^\ell_{ab}$ and $\mu$ term are of order TeV, so they may give important effects. 

We assume that the LFV decay $h\rightarrow \tau \mu$ occurs at one-loop level, and the decay rate can be written as $\Gamma(h\to\tau\mu) = \Gamma(h\to\bar{\tau}\mu) + \Gamma(h\to\tau\bar{\mu})$ \cite{Arganda:2004bz}, where 
\bea
\Gamma(h\to\tau\mu)  &=& \frac{1}{16\pi M_{h}} \left [ \left( 1- \left( \frac{M_{\tau}+M_{\mu}}{M_h} \right)^2 \right) \left( 1- \left( \frac{M_{\tau}-M_{\mu}}{M_h} \right)^2 \right) \right]^{1/2}\nonumber\\
&\times& \left[ \left(M^2_h - M^2_{\tau} - M^2_{\mu}\right) \left(|F_L|^2+|F_R|^2\right) -4 M_{\tau}M_{\mu}~ {\rm Re}(F_LF^\ast_R) \right],
\eea
and $F_L = F^{\chi^+\chi^-\tilde{\nu}}_L + F^{\chi^0\tilde{\ell}\tilde{\ell}}_L+ F^{\tilde{\nu}\tilde{\nu}\chi^\pm}_L +F^{\chi^0\chi^0\tilde{\ell}}_L$ with 

\begin{equation}\hspace{-6.1cm}
F^{\chi^0\tilde{\ell}\tilde{\ell}}_L = - \left[ \Gamma^{\bar{\chi}^0 \ell \tilde{\ell}^\ast}_L \Gamma^{\bar{\chi}^0 \bar{\ell} \tilde{\ell}}_L \Gamma^{h \tilde{\ell} \tilde{\ell}^\ast} \right] M_{\chi^0_i} ~C_{0}(M_{\tilde{\chi}^{0}}^{2}M_{\tilde{l}_{i}}^{2},M_{\tilde{l}_{j}}^{2}),
\end{equation}

\begin{equation}\hspace{-5.4cm}
F^{\chi^\pm\tilde{\nu}\tilde{\nu}}_L = - \left[ \Gamma^{\bar{\chi}^+ \ell \tilde{\nu}^\ast}_L \Gamma^{\bar{\chi}^- \bar{\ell} \tilde{\nu}}_L \Gamma^{h \tilde{\nu} \tilde{\nu}^\ast} \right] M_{\chi^\pm} ~C_{0}(M_{\tilde{\chi}^{\pm}}^{2}M_{\tilde{\nu}_{i}}^{2},M_{\tilde{\nu}_{j}}^{2}),
\end{equation}

\begin{equation*}\hspace{-3.8cm}
F^{\chi^\pm \chi^\pm \tilde{\nu}}_L = - \left[ \Gamma^{\bar{\chi}^+ l \tilde{\nu}^\ast}_L \Gamma^{\bar{\chi}^- \bar{l} \tilde{\nu}}_L \left(\Gamma^{h \tilde{\chi}^+ \tilde{\chi}^-}_LM_{\chi^\pm_a}M_{\chi^\pm_b}~C_{0}(M_{\tilde{\chi}^{\pm}_{a}}^{2},M_{\tilde{\chi}^{\pm}_{b}}^{2},M_{\tilde{\nu}_{i}}^{2})\right. \right.
\end{equation*}

\begin{equation}
\left. \left. +\Gamma^{h \tilde{\chi}^+ \tilde{\chi}^-}_R\left(B_{0}(0,M_{\tilde{\chi}^{\pm}_{a}}^{2},M_{\tilde{\chi}^{\pm}_{b}}^{2})+M_{\tilde{\nu}_i}^{2}C_{0}(M_{\tilde{\chi}^{\pm}_{a}}^{2},M_{\tilde{\chi}^{\pm}_{b}}^{2},M_{\tilde{\nu}_{i}}^{2})\right)\right) \right],
\end{equation}

\begin{equation*}\hspace{-4.6cm}
F^{\chi^0\chi^0\tilde{\ell}}_L = - \left[ \Gamma^{\bar{\chi}^0 \ell \tilde{\ell}^\ast}_L \Gamma^{\bar{\chi}^0 \bar{\ell} \tilde{\ell}}_L \left(\Gamma^{h \tilde{\chi}^0 \tilde{\chi}^0}_LM_{\chi^0_a}M_{\chi^0_b}C_{0}(M_{\tilde{\chi}^{0}_{a}}^{2},M_{\tilde{\chi}^{0}_{b}}^{2},M_{\tilde{l}_{i}}^{2})\right. \right. 
\end{equation*}

\begin{equation}\hspace{-1.0cm}
\left. \left. +\Gamma^{h \tilde{\chi}^0 \tilde{\chi}^0}_R(B_{0}(0,M_{\tilde{\chi}^{0}_{a}}^{2},M_{\tilde{\chi}^{0}_{b}}^{2})+M_{\tilde{l}_i}^{2}C_{0}(M_{\tilde{\chi}^{0}_{a}}^{2},M_{\tilde{\chi}^{0}_{b}}^{2},M_{\tilde{l}_{i}}^{2}))\right) \right].
\end{equation}
Here we neglect the tau and muon masses. The expression of $F_R$ can be obtained from $F_L$ by exchanging $L \to R$. Note that the form factors given above are enhanced with the masses of supersymmetric particles, in contrast to  the Loop functions which are given by \cite{Hollik:1988ii}

\begin{equation}
B_{0}(0,x,y) = 1-\log \frac{x}{\mu^{2}}+\frac{x \log\dfrac{y}{x}}{x-y}
\end{equation}

\begin{equation}\hspace{1.2cm}
C_{0}(x,y,z) = \frac{1}{y-z}\left(\frac{z\log \dfrac{z}{x}}{x-z}+ \frac{y\log \dfrac{y}{x}}{y-x}\right)
\end{equation}

If we consider the basis in which the charged lepton mass matrix is diagonal, i.e. $Y_{ij}^{l}=(m_{ii}^{l}/v)\delta_{ij}$, then the terms proprtional to $Y_{ij}^{l}$ in the above expressions are quite suppressed. On the other hand, the vertices $\Gamma^{\chi^+_j \ell_i \tilde{\nu}_k}_R$,  $\Gamma^{\chi^0_j \ell_i \tilde{\ell}_k}_L $ and $\Gamma^{\chi_k^0 \ell_i \tilde{\ell}_k}_R $ can be enhanced with non-universal slepton mass matrix and bino like neutralino. In addition, as mentioned in the previous section, the trilinear scalar interaction coupling $T_{\ell}$ and bilinear Higgs mixing term $\mu$ may enhance the relevant couplings of Higgs. However, in constrained MSSM (CMSSM), where universal SSB terms are imposed at $M_{{\rm GUT}}$, the off-diagonal terms in the slepton/sneutrino mass matrices, induced through RGEs, are quite negligible. In addition, due to the absence of right-handed neutrinos, one can adopt a diagonal basis of charged mass matrix in which $V^{l}_{L,R}=\mathbb{1}$. Therefore, the slepton/sneutrino mass matrices remain almost diagonal in the family basis at the electroweak scale. As a result, CMSSM does not provide any source for LFV processes which can be probed by the signal. 

However, in a generic framework of MSSM, one can assume non-universal SSB terms at either the GUT scale or the electroweak scale. In such a non-universal setup for MSSM, one can identify two sources which contribute to the LFV decay $h\rightarrow \tau \mu$. One includes the off-diagonal elements in the slepton/sneutrino mass matrices, and the other is from the off-diagonal elements in the trilinear scalar interaction coupling. Note that in the latter case, $T_{l}$ is not factorized in terms of $Y_{l}$. While these two sources contribute to the Higgs boson decay into $\tau$ and $\mu$, they also contribute to other LFV decays such as $\tau \rightarrow \mu \gamma$, $\tau \rightarrow e \gamma$, and $\mu \rightarrow e \gamma$; and hence, the contributions from these sources can be highly constrained by the experimental bounds on these decay processes. 

We perform two different scans over the following parameters to analyze the contributions from the sources mentioned above separately:

\begin{itemize}
\item Diagonal $T-$term: 

$0 \leq m_{\tilde{\mu}_{L,R}},m_{\tilde{\tau}_{L,R}},m_{LL}(2,3),M_{i} \leq 5 $ TeV, 

$0\leq \tan\beta \leq 60$, $|T_{l}|\leq 15$ TeV.

\item Diagonal slepton mass matrix: 

$0 \leq m_{\tilde{\mu}_{L,R}},m_{\tilde{\tau}_{L,R}},M_{i} \leq 5 $ TeV,

$|T_{l}(2,3)|,|T_{l}(3,2)|\leq 15$ TeV, $0\leq \tan\beta \leq 60$.
\end{itemize}
where $m_{\tilde{\mu}_{L,R}}, m_{\tilde{\tau}_{L,R}}$ are the SSB masses of smuon and stau, while $m_{LL}(2,3)$ stands for the off-diagonal element of slepton mass matrix, which mixes the smuon and stau. $M_{i}$ ($i=1,2,3$) are the SSB gaugino mass terms for $U(1)_{Y}$, $SU(2)_{L}$ and $SU(3)_{c}$ respectively. $T_{l}$ is the trilinear scalar interaction coupling, and $\tan\beta$ is the ratio of vacuum expectation values (VEVs) of the MSSM Higgs doublets. In these scans, we employ SPheno \cite{Porod:2011nf} obtained by using SARAH \cite{Staub:2013tta}

In the first set, we keep $T_{l}$ diagonal and vary $m_{L}(2,3)$, which is the corresponding off-diagonal element for $\mu - \tau$ mixing in the slepton/sneutrino mass matrix. Even though we restrict the scan over the mass matrix of the left-handed sleptons, the similar discussion holds for the case in which the mixing happens in the right-handed sleptons. On the other hand, a less enhancement should be expected when the family mixing is placed in the right-handed slepton sector, since $SU(2)$ interactions do not contribute due to the chirality in this case. Similarly, we keep the mass matrix diagonal in the second scan, while we vary the off-diagonal elements of $T_{l}$. We neglected the terms in these matrices, which do not contribute to $h\rightarrow \tau \mu$ for simplicity. In this case, the bounds for $\tau \rightarrow e \gamma$ and $\mu \rightarrow e \gamma$ can easily be satisfied, while $\tau \rightarrow \mu \gamma$ still puts a severe bound on the results. 

\begin{figure}[ht!]
\centering
\includegraphics[scale = 1.0]{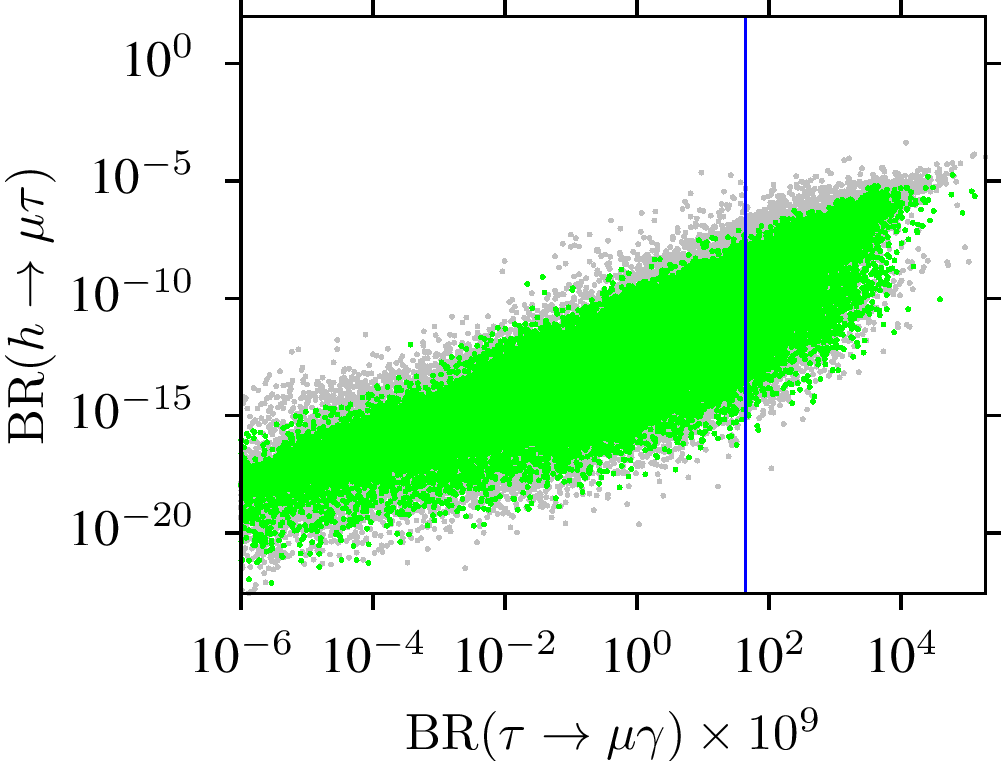}~~~~
\includegraphics[scale = 1.0]{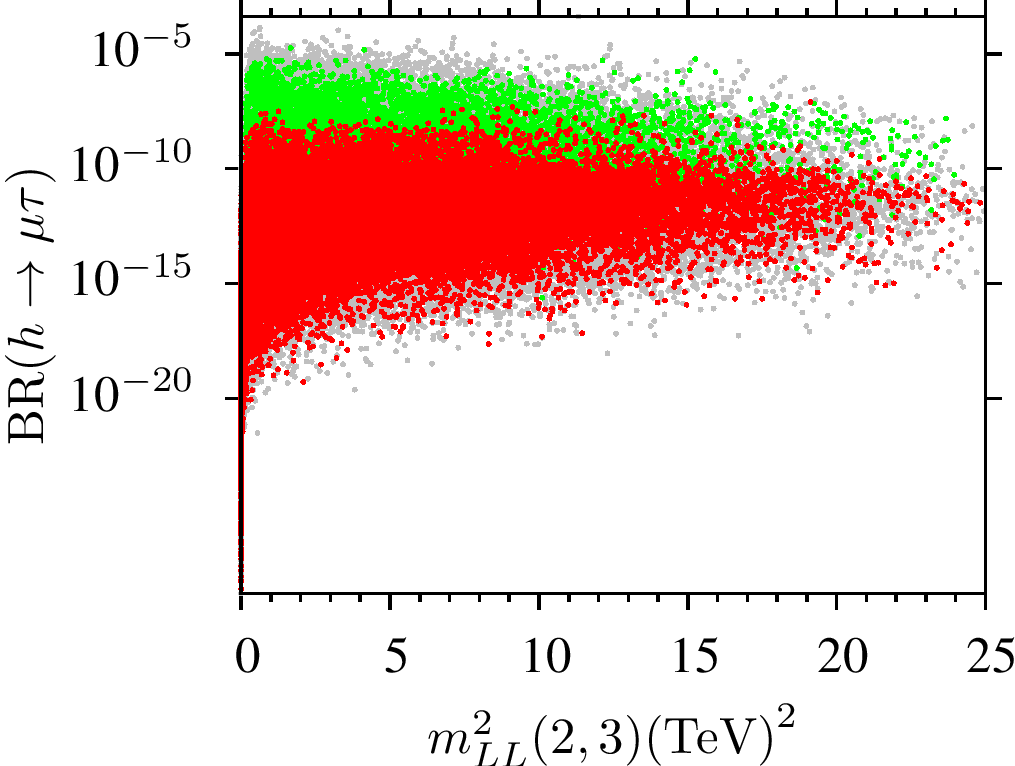}
\caption{(Left) Correlation between $BR(h \to \tau \mu)$ and $BR(\tau \to \mu \gamma )$ in the MSSM. (Right) $BR(h \to \tau \mu)$ versus the slepton off-diagonal mass term $(m^{\tilde{\ell}}_{LL})_{23}$ in the MSSM with non-diagonal slepton mass matrix as given in Eq.\ref{mLL2}. While gray points are excluded by the LHC constraints, green points satisfy the mass bounds on sparticles and the constraints from the rare B-meson decays. In addition, the vertical line in the left panel indicates the bound on $BR(\tau \rightarrow \mu \gamma)$, and the red points in the right panel form a subset of green and they satisfy the bound on $BR(\tau \rightarrow \mu \gamma)$.}
\label{mL23}
\end{figure} 

\begin{figure}[h!]
\centering
\includegraphics[scale = 1.0]{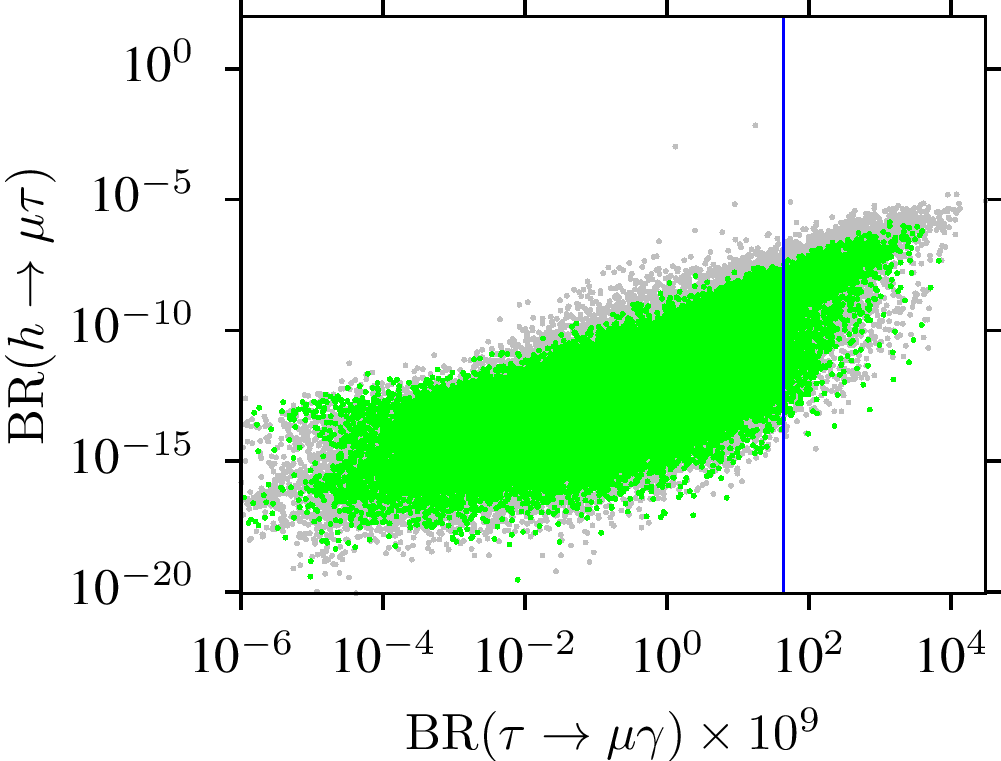}~~~~
\includegraphics[scale = 1.0]{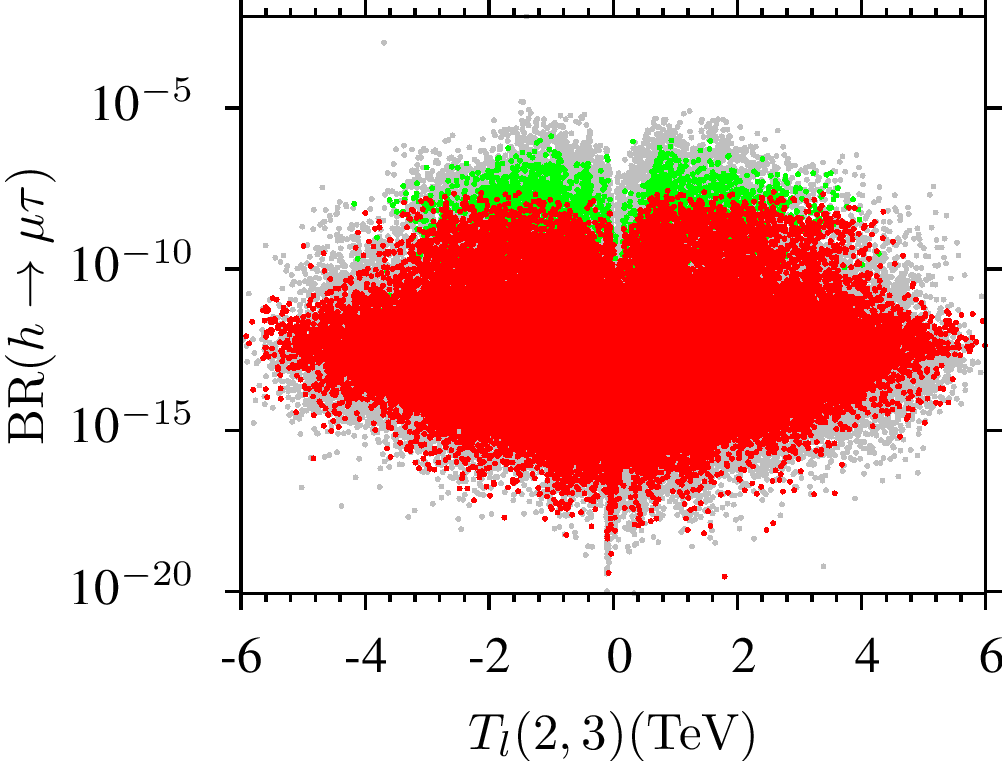}
\caption{(Left) Correlation between $BR(h \to \tau \mu)$ and $BR(\tau \to \mu \gamma )$. (Right) $BR(h \to \tau \mu)$ versus the off-diagonal element of trilinear coupling $T^{\tilde{\ell}}_{23}$ in the MSSM with non-universal trilinear couplings. The color coding is the same as Figure \ref{mL23}.}
\label{TE23}
\end{figure}

Figure \ref{mL23} displays the results from the scan with non-diagonal slepton mass matrix as given in Eq.\ref{mLL2} in terms of correlations between $BR(h \to \tau \mu)$ and $BR(\tau \to \mu \gamma )$, and  also $BR(h \to \tau \mu)$ versus the slepton off-diagonal mass term $(m^{\tilde{\ell}}_{LL})_{23}$. While gray points are excluded by the LHC constraints, green points satisfy the mass bounds on sparticles and the constraints from the rare B-meson decays. In addition, the vertical line in the left panel indicates the bound on $BR(\tau \rightarrow \mu \gamma)$, and the red points in the right panel form a subset of green and they satisfy the bound on $BR(\tau \rightarrow \mu \gamma)$. As seen from the left panel, $BR(h \to \tau \mu)$ can be as large as about $10^{-5}$ which is three magnitudes smaller than the values that can be probed, but this region violates the bound on $BR(\tau \to \mu \gamma )$. The maximum value for $BR(h \to \tau \mu)$ is about $10^{-8}$ without violating the bound on $BR(\tau \to \mu \gamma )$. 

Similar discussion can be followed for the results obtained for the MSSM with non-universal trilinear couplings, as shown in Figure \ref{TE23}. The color coding is the same as Figure \ref{mL23}. In conclusion, even though one can scan over a wider range of the parameters, it is not possible to enhance $BR(h \to \tau \mu)$ without violating the experimental bound on $BR(\tau \to \mu \gamma )$. Our results are consistent with Ref. \cite{Aloni:2015wvn}, where it is confirmed that a large $BR(h\rightarrow \tau \mu)$ in MSSM is not possible.

\section{$h\to \tau \mu$ in Supersymmetric Seesaw Model}
\label{sec:SS1}

Seesaw mechanism is an elegant way to generate very light neutrino
masses by introducing  heavy SM singlets (right-handed Majorana neutrinos). 
In SSM with three families of right-handed neutrino superfields $N^c_i, i=1,2,3$, the superpotential for the lepton 
section is given by \cite{Casas:2001sr}
\be 
W = (Y_\ell)_{ij} E^c_i L_j H_d + (Y_\nu)_{ij} N^c_i L_j H_u + \frac{1}{2} (M_R)_{ij} N^c_i N^c_j , 
\ee
where $i,j$ run over generations, and $M_{R}$ is the heavy right-handed neutrino mass matrix. 
After the electroweak symmetry breaking, one finds  the following neutrino mass
\be 
m_\nu = m_D^T M_R^{-1} m_D
\ee
where $m_D$ is the Dirac mass matrix, given by $m_D = Y_\nu \langle H_u \rangle = v\sin \beta Y_\nu$. 
The symmetric light neutrino mass matrix $m_\nu$ can be diagonalized by a unitary matrix $U$: $U^T m_\nu U= {\rm diag}\left(m_{\nu_1}, m_{\nu_2},m_{\nu_3}\right)$. In general the Maki-Nakagawa- Sakata (MNS) mixing matrix, $U_{\rm MNS}$, is given by $U_{\rm MNS}=U U_\ell^+$, where $U_\ell$ is the charged lepton mixing matrix. In the basis of diagonal charged lepton, $U= U_{\rm MNS}$.  One of the interesting parametrization for the Dirac neutrino mass matrix is given
by
\be
m_D = U_{\rm MNS} \sqrt{m_\nu^{\rm diag}} R \sqrt{M_R}.
\label{ibara}
\ee
where $m_\nu^{\rm diag}$ is the physical light neutrino mass matrix, $U_{\rm MNS}$ is the lepton mixing
matrix. The matrix $R$ is an arbitrary orthogonal matrix which can be in general parameterized in terms of three complex angles. 

In SSM, off-diagonal elements of $m_{\tilde{L}}^2$ are naturally induced through the radiative corrections. This can be seen from the renormalization group equations of $m_{\tilde{L}}^2$ and $T_{\ell}$, which are now given by 
\bea
\frac{d m_{\tilde{L}}^2}{d t} &=& \left(\frac{d m_{\tilde{L}}^2}{d t}\right)_{\rm MSSM} + \frac{1}{16 \pi^2} \left\{ (m_{\tilde{L}}^2 Y_\nu^+ Y_\nu) +   
(Y_\nu^+ Y_\nu m_{\tilde{L}}^2 )  + 2 (Y_\nu^+ Y_\nu ) m_{H}^2 \right.\nonumber\\\nonumber \\
&+& \left.  (Y_\nu^+ m_{\tilde{\nu}}^2 Y_\nu ) + 2 (T_\nu^\dag T_\nu)\right\},
\eea

\begin{equation}\hspace{-4.6cm}
\frac{d T_\ell}{dt} = \left(\frac{d T_\ell}{dt} \right)_{MSSM} + 2 Y_\ell Y_\nu^\dag T_\nu + T_\ell Y_\nu^\dag Y_\nu. 
\label{TRGE}
\end{equation}

Therefore, the off-diagonal elements of $m_{\tilde{L}}^2$ can be generated by non-diagonal $Y_\nu$ even if the soft terms $m_{\tilde{L}}$ and $T_{l}=Y_{l}A_{0}$ are universal at GUT scale. On the other hand, the non-diagonal entries radiatively generated through Eq.(\ref{TRGE}) are quite negligible, since $Y_{\ell}$ is also assumed to be diagonal. The family mixing in slepton sector generated through RGEs yields non-zero contributions to $\mu\rightarrow e \gamma$ which gives a strict constraint on the results.  

We have performed random scan over the following parameter space 

\begin{equation}
\begin{array}{ccc}
0 \leq & m_{0} & \leq 3~(\rm TeV) \\
0 \leq & M_{1/2} & \leq 3~(\rm TeV) \\
-3 \leq & A_{0}/m_{0} & \leq 3 \\
1.2 \leq & \tan\beta &  \leq 60 \\
10^{7} \leq & M_{R_{1}},M_{R_{2}},M_{R_{3}} & \hspace{0.5cm}\leq 10^{14}~(\rm GeV)
\end{array}
\end{equation}
and we require the solutions to have an orthogonal matrix $R$ which yields light neutrino masses consistently with the experimental results accordingly to Eq.(\ref{ibara}). 

\begin{figure}[ht!]
\centering
\includegraphics[scale = 1.0]{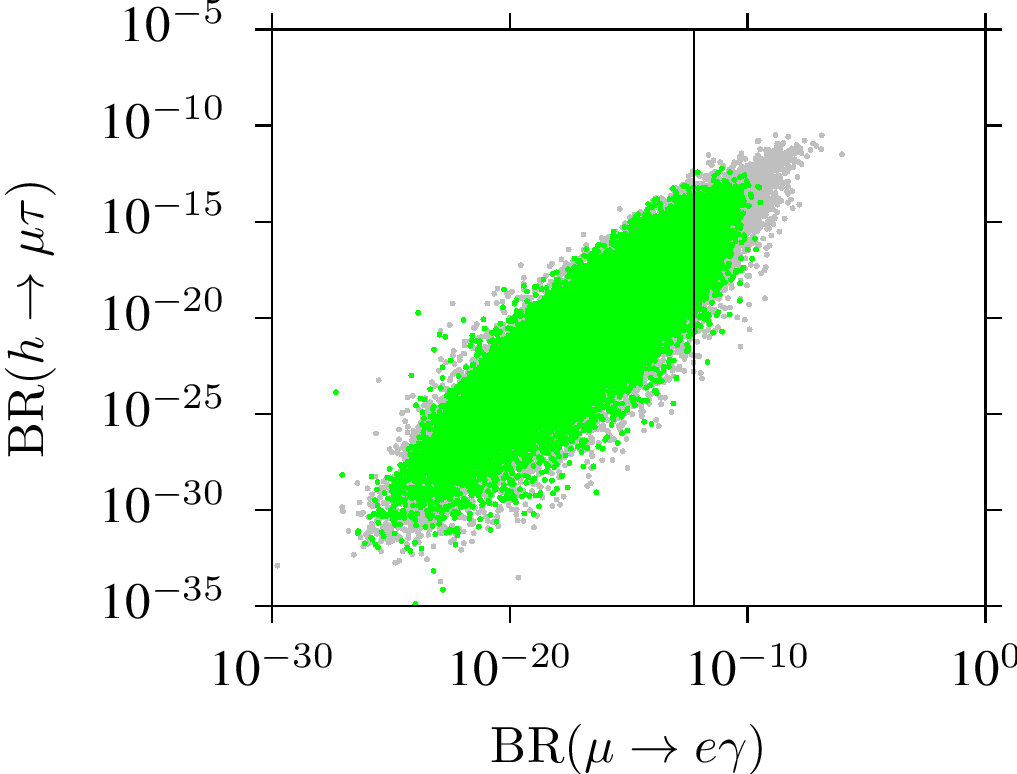}~~ \includegraphics[scale = 1.0]{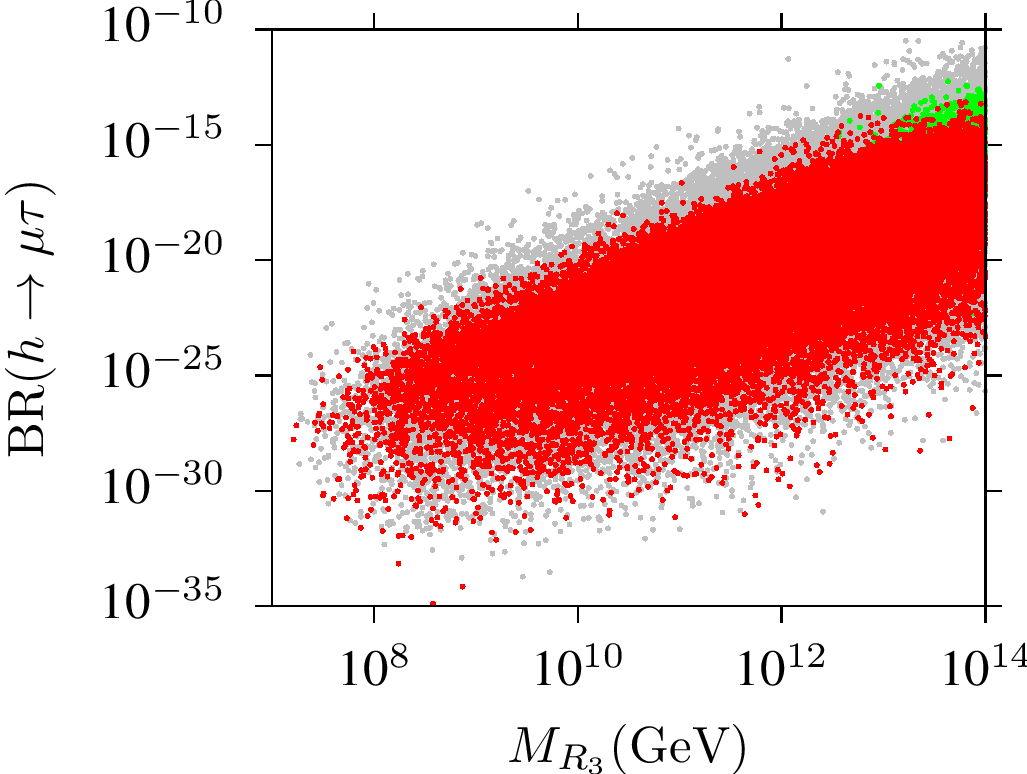}
\caption{(Left) Correlation between $BR(h\rightarrow \tau \mu)$ and $BR(\mu\rightarrow e \gamma)$. (Right) Correlation between $BR(h\rightarrow \tau \mu)$ and $M_{R_{3}}$. The color coding is the same as Figure \ref{mL23} except the vertical line in the left panel indicates the bound on $BR(\mu\rightarrow e \gamma)$, and the red points satisfy the experimental bound on $BR(\mu\rightarrow e \gamma)$. }
\label{SSM1}
\end{figure}

We found that the bounds on the LFV decays $\tau\rightarrow \mu \gamma$ and $\tau \rightarrow e \gamma$ can easily be satisfied, while $\mu \rightarrow e \gamma$ provides a severe constraint on the enhancement in the $BR(h\rightarrow \tau \mu)$ as seen in Figure \ref{SSM1}. We display our results in correlations between $BR(h\rightarrow \tau \mu)$ and $BR(\mu\rightarrow e \gamma)$ (left), $M_{R_{3}}$ (right). The color coding is the same as Figure \ref{mL23} except the vertical line in the left panel indicates the bound on $BR(\mu\rightarrow e \gamma)$, and the red points satisfy the experimental bound on $BR(\mu\rightarrow e \gamma)$. The left panel shows that $BR(h\rightarrow \tau \mu)$ can only be as large as about $10^{-15}$ without violating the bound on the $BR(\mu\rightarrow e \gamma)$. The right panel reveals a linear correlation between $BR(h\rightarrow \tau \mu)$ and $M_{R_{3}}$. This is because $M_{R_{3}}$ is effective to induce the off-diagonal elements which mixes $\tau$ and $\mu$, and according to the results shown in the right panel of Figure \ref{SSM1}, a significant enhancement requires very heavy $M_{R_{3}}$. The results represented in this section are consistent with those in Ref. \cite{Arganda:2004bz}, and this section updates the findings in the light of the current experimental constraints from the Higgs boson mass and LFV processes.

\section{$h\to \tau \mu$ in BLSSM with Inverse Seesaw}
\label{sec:BLSSMISS}

The minimal extension of the SM, based on the gauge group $SU(3)_C \times SU(2)_L  \times U(1)_Y \times U(1)_{B-L}$, also provides a suitable framework for the inverse seesaw mechanism, which can naturally account for light neutrino masses \cite{Khalil:2010iu}. The particle content of supersymmetric version of this model (BLSSM-IS) includes the following superfields in addition to those in MSSM: $(i)$ two SM singlet chiral Higgs superfields $\chi_{1,2}$ which are responsible for  
$U(1)_{B-L}$ breaking $(ii)$ three sets of SM singlet chiral
superfields, $\nu_i, s_{1_i}, s_{2_i} (i =1,2,3)$, to implement the
IS mechanism, without generating $B-L$ anomaly.  The Superpotential in this model is given by \cite{Elsayed:2011de}
\bea
W =  - \mu' \,\hat{\chi}_1\,\hat{\chi}_2\,+\mu\,\hat{H}_u\,\hat{H}_d\,+\mu_S\,\hat{s}_2\,\hat{s}_2\,- Y_d \,\hat{d}\,\hat{q}\,\hat{H}_d\,- Y_\ell \,\hat{e}\,\hat{l}\,\hat{H}_d 
+Y_u\,\hat{u}\,\hat{q}\,\hat{H}_u  \nonumber \\
+Y_s\,\hat{\nu}\,\hat{\chi}_1\,\hat{s}_2 +Y_\nu\,\hat{\nu}\,\hat{l}\,\hat{H}_u. \hspace{6.8cm}
\label{superpotential}
\eea

After the $B-L$ and  electroweak symmetry breaking,  one finds that the neutrinos mix with the fermionic singlet fields to build up the following $9\times 9$ mass matrix, in the basis 
$(\nu_L ,\nu^c, S_2)$:
\be {\cal M}_{\nu}=
\left(%
\begin{array}{ccc}
  0 & m_D & 0\\
  m^T_D & 0 & M_R \\
  0 & M^T_R & \mu_s\\
\end{array}%
\right), %
\label{inverse}
\ee%
where $m_D=\frac{1}{\sqrt{2}}Y_\nu v$ and $ M_R =
\frac{1}{\sqrt 2}Y_{s} v' $. The VEVs of the Higgs fields are defined as $\langle{\rm Re} H_i^0\rangle=\frac{v_i}{\sqrt{2}}$ and $\langle{\rm Re} \chi^0_i\rangle=\frac{v'_i}{\sqrt{2}}$, with $v=\sqrt{v^2_1+v^2_2}\simeq 246$ GeV and $v'=\sqrt{v'^2_1+v'^2_2}$. Moreover, one may radiatively generate a very small Majorana mass for the
$S_{2}$ fermion through possible non-renormalisable terms.  
The diagonalisation of the mass matrix, Eq. (\ref{inverse}), 
leads to the following light and heavy neutrino masses, respectively: %
\begin{eqnarray}%
m_{\nu_l} &=& m_D M_R^{-1} \mu_s (M_R^T)^{-1} m_D^T,\label{mnul}\\
m_{\nu_H}&=& m_{\nu_{H'}} = \sqrt{M_R^2 + m_D^2}. %
\end{eqnarray} %
Thus, one finds that the light neutrino masses can be of order eV, with a TeV scale $M_R$, if $\mu_s \ll M_R$, and large Yukawa coupling $Y_{\nu}\sim \mathcal{O}(1)$. Such a large coupling is a feature of the BLSSM-IS. In this case, the Dirac mass matrix can be written as  
\be
m_D = U_{\rm MNS} \sqrt{m_\nu^{\rm diag}} R  \sqrt{\mu_s^{-1}} M_R.
\label{ibara2}
\ee

We perform a random scan over the following parameter space

\begin{equation}
\begin{array}{ccc}
0 \leq & m_{0} & \leq 5~(\rm TeV) \\
0 \leq & M_{1/2} & \leq 5~(\rm TeV) \\
-3 \leq & A_{0}/m_{0} & \leq 3 \\
1.2 \leq & \tan\beta &  \leq 60 \\
1 \leq & \tan\beta' &  \leq 2 \\
\end{array}
\end{equation}
$\mu_{S} \sim 10^{-7}$ GeV and $M_{Z'}=2.5$ TeV are fixed. We also require our solutions that there is always an orthogonal matrix $R$ which yield correct neutrino masses. 

\begin{figure}[ht!]
\includegraphics[scale=1]{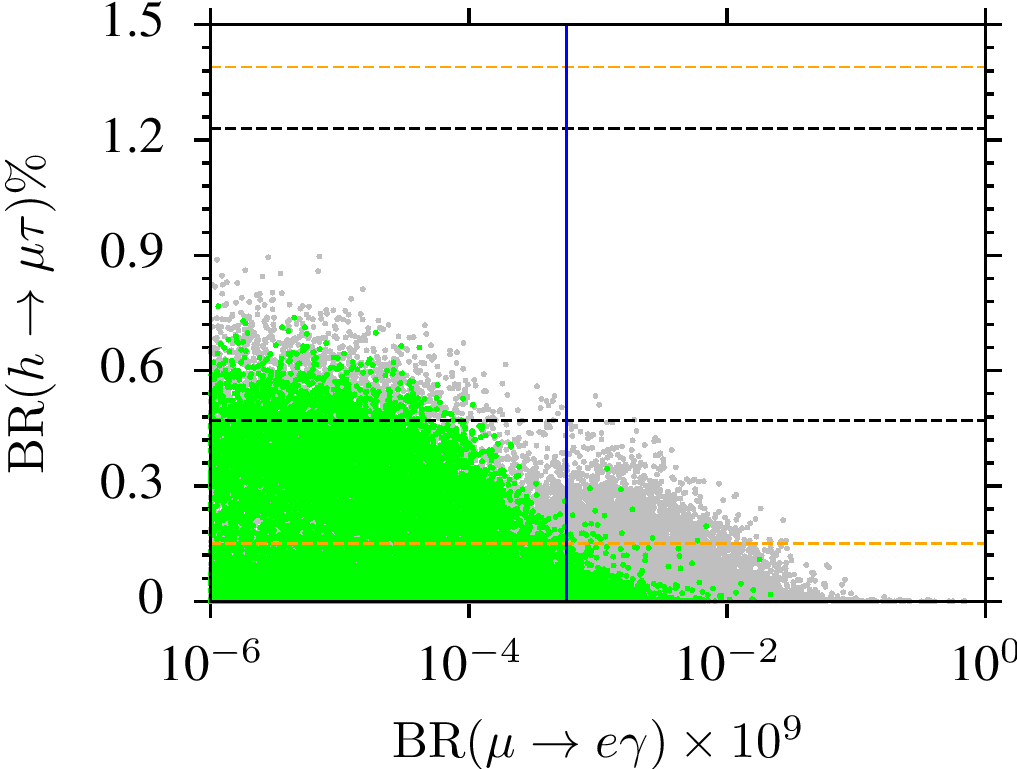}
\includegraphics[scale=1]{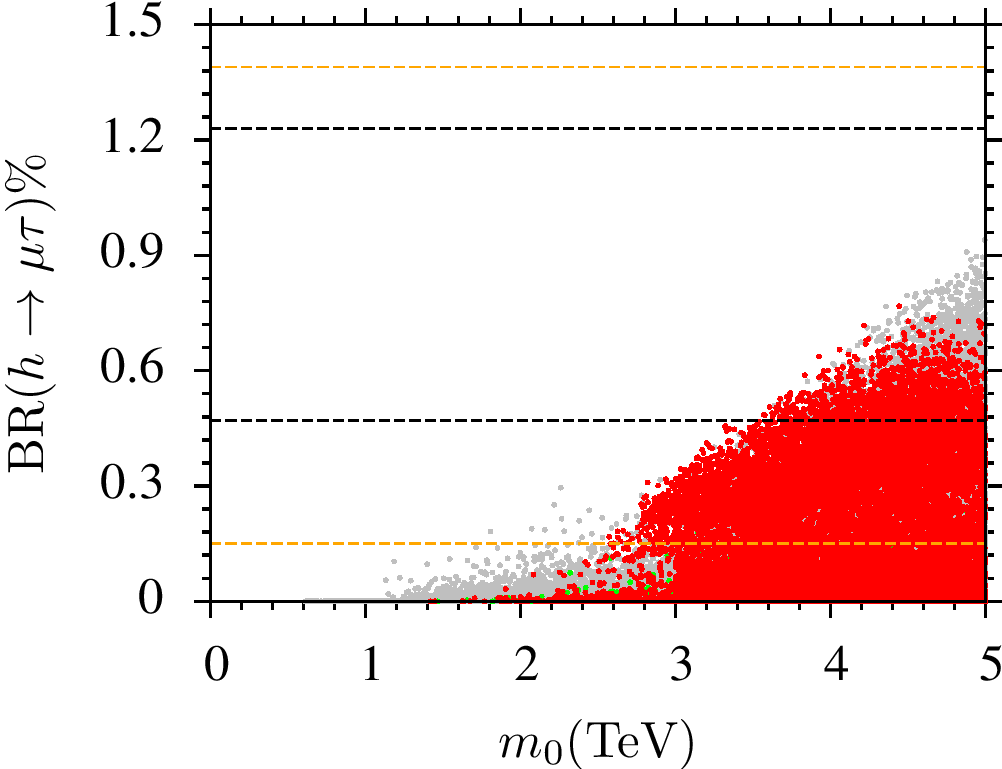}
\includegraphics[scale=1]{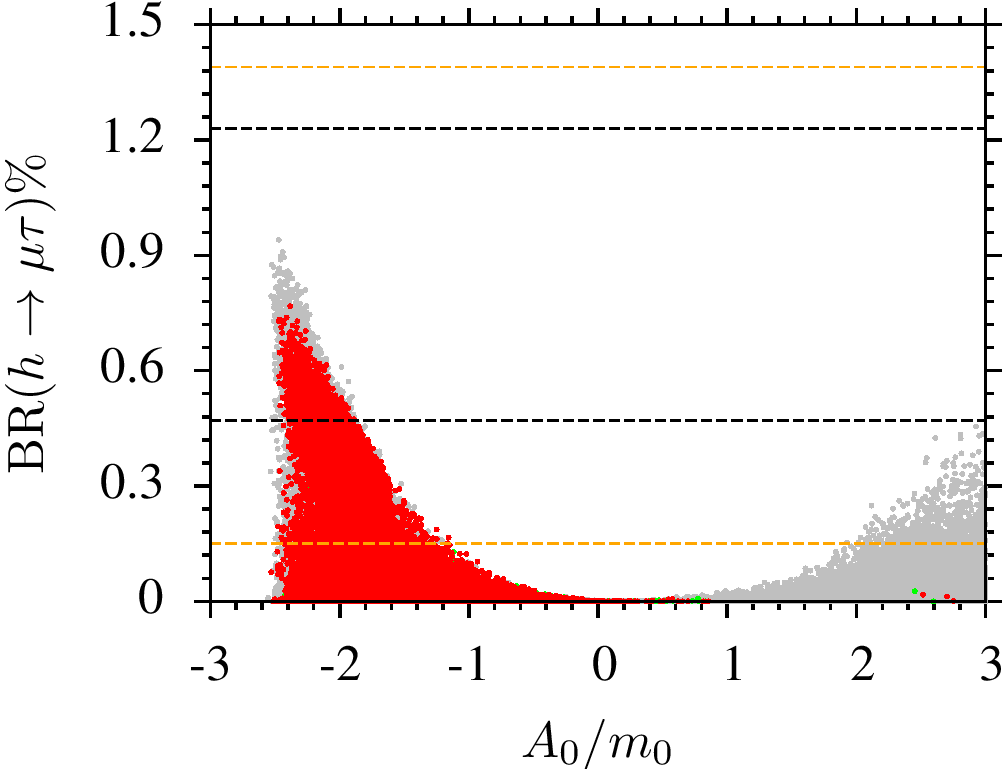}
\hspace{1.6cm}\includegraphics[scale=1]{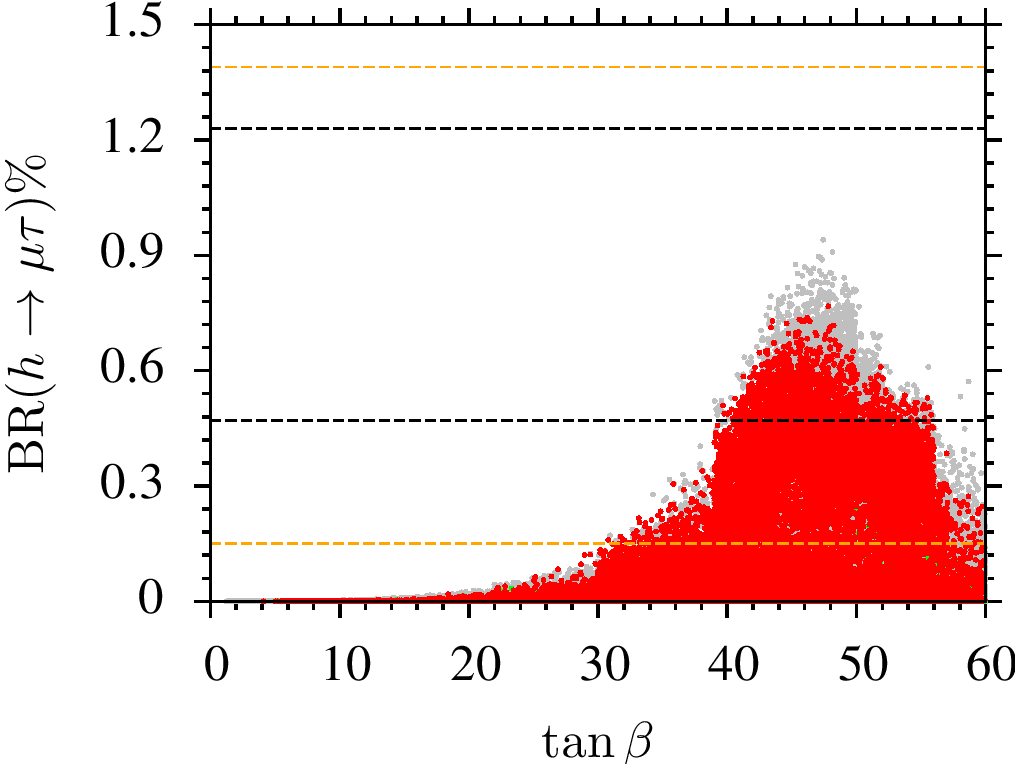}
\caption{Plots in the $BR(h\rightarrow \tau \mu) - BR(\mu\rightarrow e \gamma)$, $BR(h\rightarrow \tau \mu) - m_{0}$, $BR(h\rightarrow \tau \mu) - A_{0}/m_{0}$, and $BR(h\rightarrow \tau \mu) - \tan\beta$ planes. Gray points are excluded by the LHC constraints, while the green points are allowed. The red points form a subset of green and they are consistent with the bound on $BR(\mu \rightarrow e \gamma)$. The black (orange) dashed lines indicate the best fit for the ${\rm BR}(h\rightarrow \mu \tau)$ obtained by CMS \cite{Khachatryan:2015kon} (ATLAS \cite{Aad:2015gha}).}
\label{BLSSMIS1}
\end{figure}

In Figure \ref{BLSSMIS1} we present the results for BLSSM-IS in the $BR(h\rightarrow \tau \mu)$ versus $BR(\mu\rightarrow e \gamma)$, and $BR(h\rightarrow \tau \mu)$ as a function of $m_{0}$, $ A_{0}/m_{0}$, and $\tan\beta$. Gray points are excluded by the LHC constraints, while the green points are allowed. The red points form a subset of green and they are consistent with the bound on $BR(\mu \rightarrow e \gamma)$. The black (orange) dashed lines indicate the best fit for the ${\rm BR}(h\rightarrow \mu \tau)$ obtained by CMS \cite{Khachatryan:2015kon} (ATLAS \cite{Aad:2015gha}). As in the case of SSM, the bounds on the LFV processes $\tau \rightarrow \mu \gamma$ and $\tau \rightarrow e \gamma$ are satisfied by all the solutions without imposing any constraint for them, while the LFV decay $\mu \rightarrow e \gamma$ can exclude a small portion of the solutions as seen from the correlation between $BR(h\rightarrow \tau \mu)$ and $BR(\mu\rightarrow e \gamma)$. In contrast to the generic MSSM and SSM, $BR(h\rightarrow \tau \mu)$ does not exhibit an enhancement with the $BR(\mu\rightarrow e \gamma)$. Indeed, the region excluded by the $\mu \rightarrow e \gamma$ yields low values for the $BR(h\rightarrow \tau \mu)$. This result can be explained by the fact that the family mixing happens in the LR section of the slepton mass matrix, while a possible enhancement in $BR(\mu\rightarrow e \gamma)$ can occur in the LL (or RR) section of the slepton mass matrix. Even though the mixing in the LL section can contribute to $BR(h\rightarrow \tau \mu)$, such contributions are rather at the order of some corrections because of the heavy slepton masses resulted from the large $m_{0}$ values as seen in the $BR(h\rightarrow \tau \mu) - m_{0}$ plane. $BR(h\rightarrow \tau \mu)$ is significantly enhanced with the increasing scalar masses at the GUT scale. This correlation rather yield a heavy spectrum at the low scale. On the other hand, it also requires a large and negative $A-$term ($\sim -2.5 m_{0}$), which can significantly lower the stau masses at the low scale. The $BR(h\rightarrow \tau \mu) - \tan\beta$ plane reveals the $\tan\beta$ enhancement, and the largest branching ratio is obtained for $\tan\beta \sim 45$, then it starts decreasing.  

\begin{figure}[ht!]
\includegraphics[scale=1]{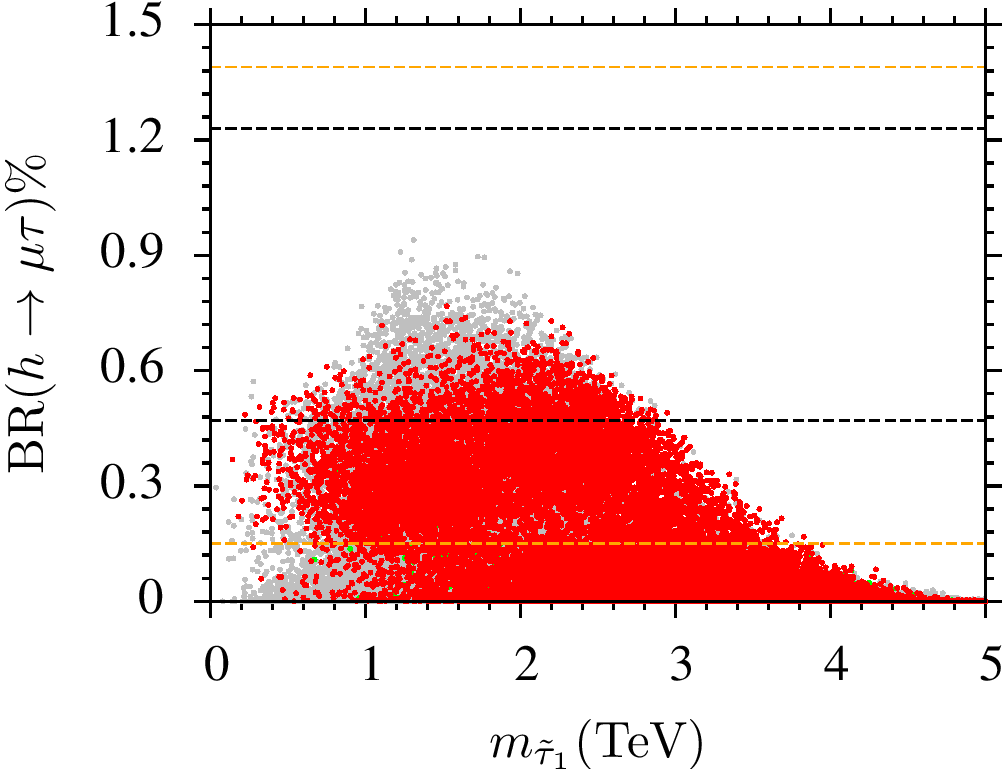}
\includegraphics[scale=1]{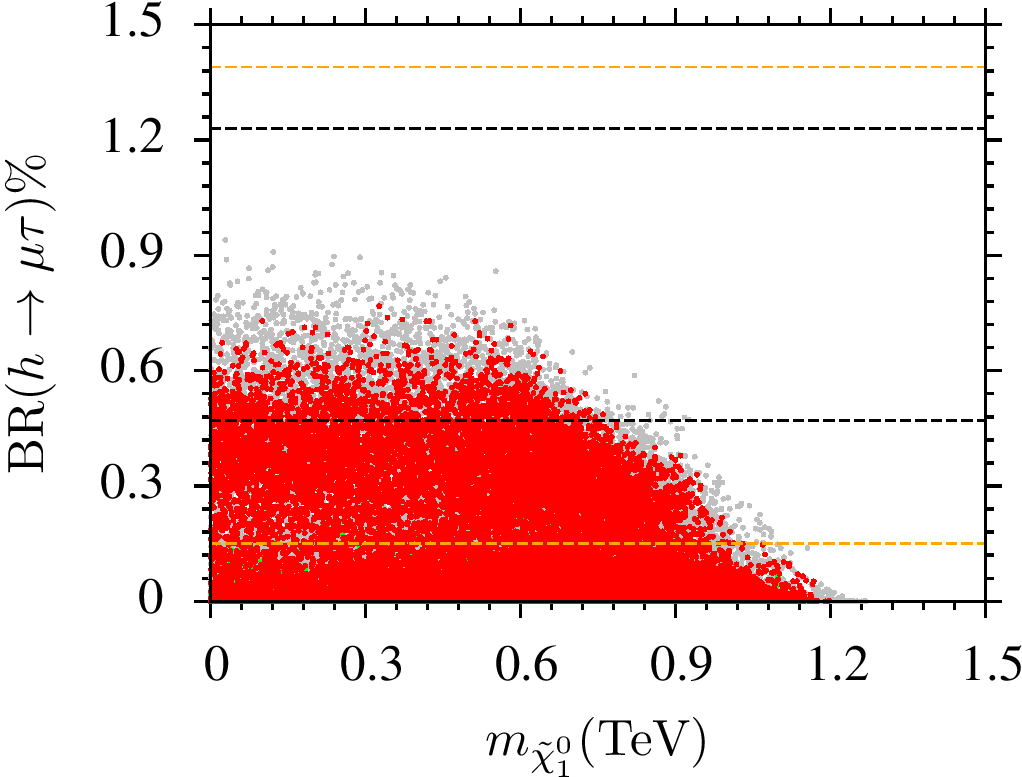}
\includegraphics[scale=1]{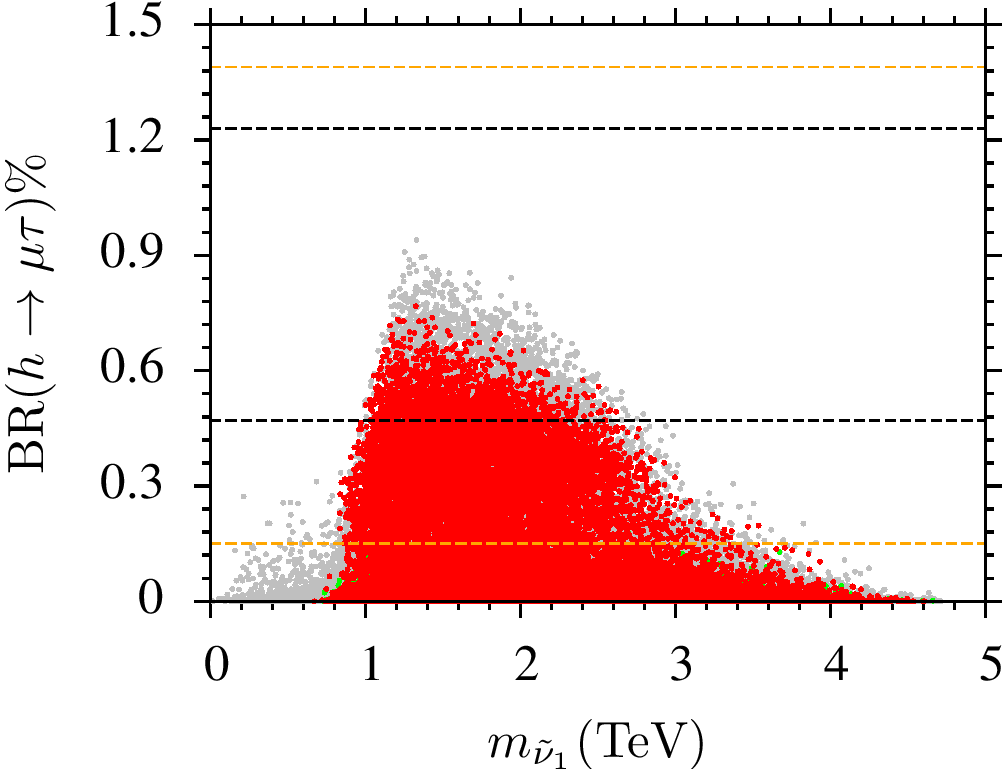}
\hspace{1.6cm}\includegraphics[scale=1]{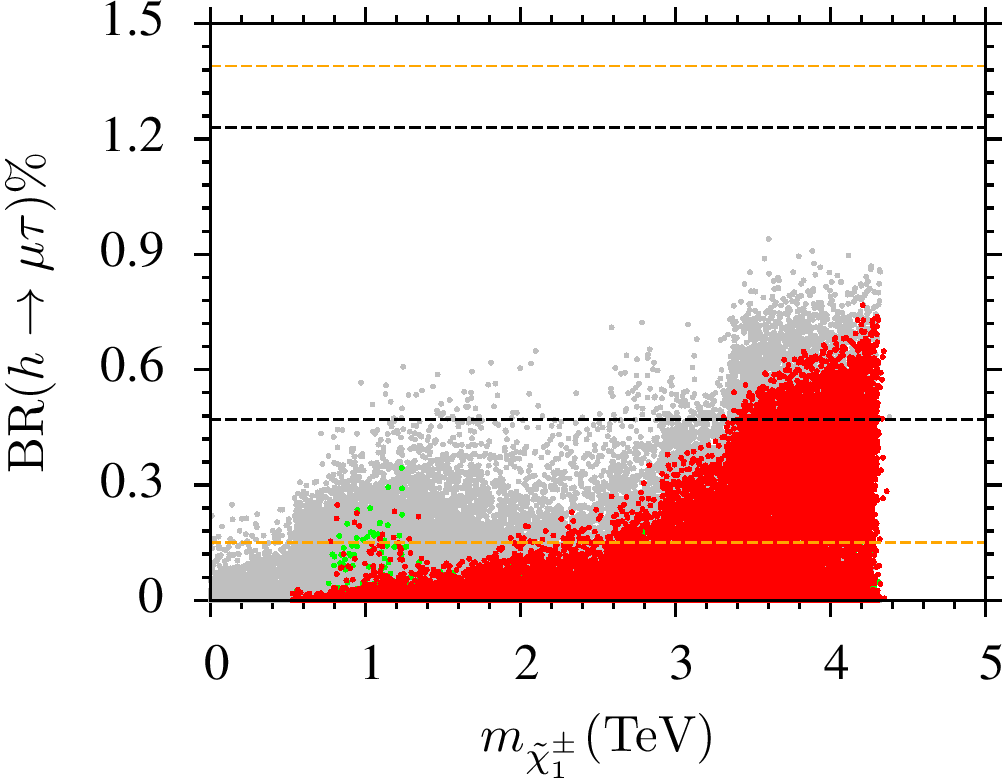}
\caption{Plots in the $BR(h\rightarrow \tau \mu) - m_{\tilde{\tau}_{1}}$, $BR(h\rightarrow \tau \mu) - m_{\tilde{\chi}_{1}^{0}}$, $BR(h\rightarrow \tau \mu) - m_{\tilde{\nu}_{1}}$, and $BR(h\rightarrow \tau \mu) - m_{\tilde{\chi}_{1}^{\pm}}$ planes. The color coding is the same as Figure \ref{BLSSMIS1}.}
\label{BLSSMIS2}
\end{figure}

Figure \ref{BLSSMIS2} displays the mass spectrum in the $BR(h\rightarrow \tau \mu) - m_{\tilde{\tau}_{1}}$, $BR(h\rightarrow \tau \mu) - m_{\tilde{\chi}_{1}^{0}}$, $BR(h\rightarrow \tau \mu) - m_{\tilde{\nu}_{1}}$, and $BR(h\rightarrow \tau \mu) - m_{\tilde{\chi}_{1}^{\pm}}$ planes. The color coding is the same as Figure \ref{BLSSMIS1}. The effective form factors, which yield to LFV Higgs decays, are enhanced by the heavy masses as mentioned in Section \ref{sec:MSSM}. On the other hand, since these supersymmetric particles are running in the loops, their masses also suppress the contributions. The compensation between the enhancement in the form factors, and suppression in the propagators in the loop requires some relatively lighter supersymmetric particles. The smuons are rather heavy because of the large $m_{0}$ mentioned above; on the other hand, stau is required to be lighter ($\lesssim 3$ TeV) as seen from the $BR(h\rightarrow \tau \mu) - m_{\tilde{\tau}_{1}}$. Similarly, the $BR(h\rightarrow \tau \mu) - m_{\tilde{\chi}_{1}^{0}}$ plane shows that the lightest neutralino mass is bounded from above at about 0.8 TeV. These two sparticles, stau and neutralino are relevant to the contribution from the first diagram given in Figure \ref{feynman}. The sneutrino mass is found to be lighter than about 3 TeV in order to have a sizable contribution from the chargino-sneutrino loop, while the chargino is much heavier ($\gtrsim 4$ TeV) in the same region. Recall that the some terms in the form factors are proportional to the $m_{\tilde{\chi}_{1}^{\pm}}^{2}$, and this explains the enhancement in $BR(h\rightarrow \tau \mu)$ with heavy chargino masses. On the other hand, one should not expect a considerable contribution from the process in which two charginos run in the loop, since the loop suppression by the heavy chargino mass takes over the enhancement in the form factors. 

\begin{figure}[ht!]
\includegraphics[scale=1]{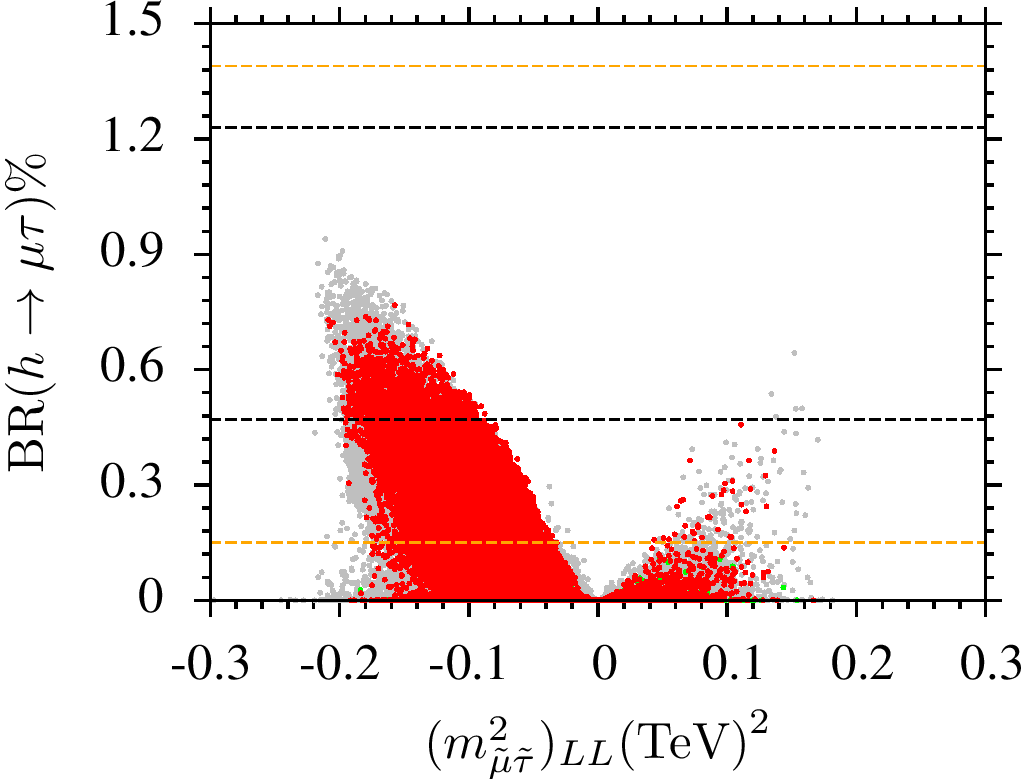}
\hspace{1.6cm}\includegraphics[scale=1]{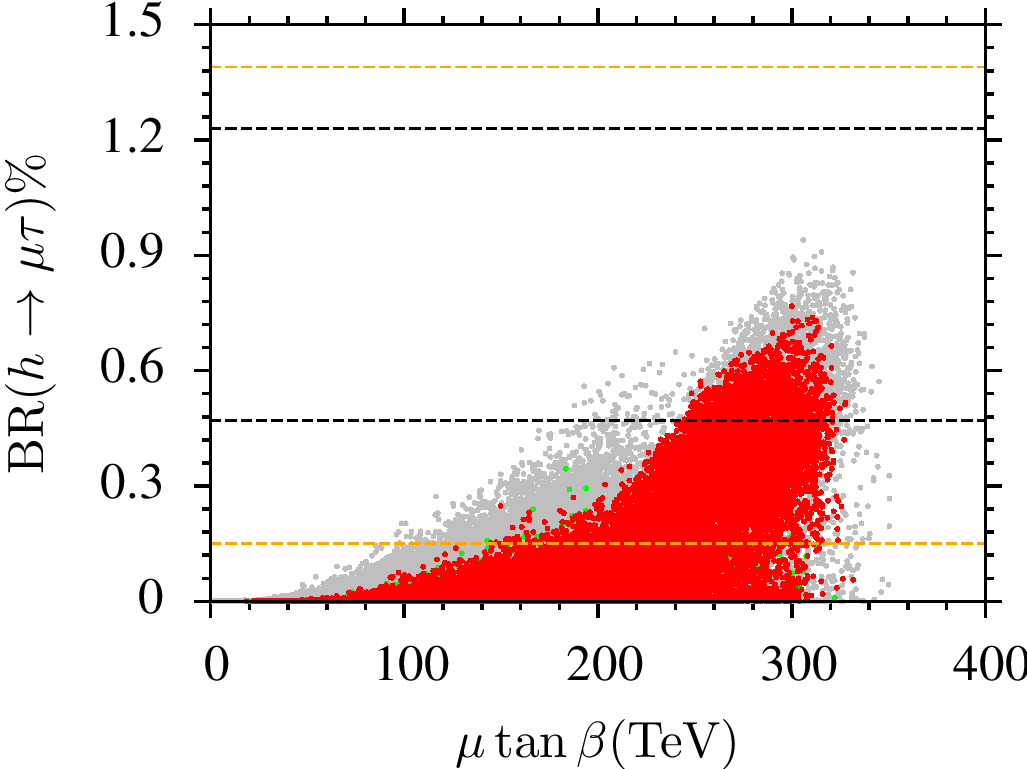}
\caption{Plots in $BR(h\rightarrow \tau \mu) -(m^{2}_{\tilde{\mu}\tilde{\tau}})_{LL}$ and $BR(h\rightarrow \tau \mu)-\mu \tan\beta$ planes. The color coding is the same as Figure \ref{BLSSMIS1}.}
\label{BLSSMIS3}
\end{figure}

Even though most of the solutions yield heavy smuon and chargino, some solutions can still be identified in the relatively light smuon ($\sim 3.5$ TeV) and light chargino ($\sim 1$ TeV) regions. These regions can provide some enhancement in $BR(h\rightarrow \tau \mu)$ through the mixing in the left-handed slepton mass matrix. The corresponding component $(m^{2}_{\tilde{\mu}\tilde{\tau}})_{LL}$ can be as large as $0.1 ({\rm TeV})^{2}$ as seen from the $BR(h\rightarrow \tau \mu) -(m^{2}_{\tilde{\mu}\tilde{\tau}})_{LL}$ plane of Figure \ref{BLSSMIS3}. On the other hand, the enhancement in this region is slightly above the lower bound of ATLAS. The heavy smuon masses in the region, which leads to the largest enhancement, can be explained with the mixing in the LR section of the slepton mass matrix, as stated above. The Higgs coupling to a left-handed slepton and right-handed slepton is enhanced with $A-$term and $\mu$, and they compensate the suppression from the heavy smuon mass. In addition, the chirality flip takes a part in such processes. As shown in the $BR(h\rightarrow \tau \mu)-\mu \tan\beta$ plane, $BR(h\rightarrow \tau \mu)$ is proportionally enhanced with the factor $\mu \tan\beta$. As a result, one can realize a significant enhancement in $BR(h\rightarrow \tau \mu)$ without violating the experimental constraints from the other LFV processes including $\mu \rightarrow e \gamma$.

\section{Search for LFV BLSSM-IS Higgs Decay at the LHC }
\label{sec:signal}

In this section we consider how likely it is to detect the LFV Higgs boson decay $h\rightarrow \tau \mu$ over the relevant SM background. 
Our selection of the final state particles is based on the analysis carried out by ATLAS \cite{Aad:2015gha}, where the selection requires the reconstruction of final state particles containing an energetic $\mu$ and $\tau$ leptons with opposite charges. Then, the background for the LFV Higgs decay can be classified into two main categories:

\begin{itemize}
\item[1.] Events with true $\tau$, and the $\mu$ lepton can be softly radiated or faked from the jets or leptons. Irreducible background processes in this category are $Z\rightarrow \tau \tau$ and $h\rightarrow \tau \tau$ with $\mu$ coming from $VV\rightarrow \tau \mu + X$, where $V$ denotes the SM gauge bosons. In addition, one of $\tau$ leptons can decay into $\mu$ in such decays \cite{Chatrchyan:2014nva,Celis:2013xja}

\item[2.] Events with fake $\tau$ signature dominated by $W+{\rm jets}$ events with some contributions from multijet, diboson VV, $t\bar{t}$, and single top events with some charge asymmetry. Such processes are reducible, and they can be eliminated by imposing veto against to jets.
\end{itemize}

\begin{figure}[ht!]
\centering
\includegraphics[width=8cm,height=5cm]{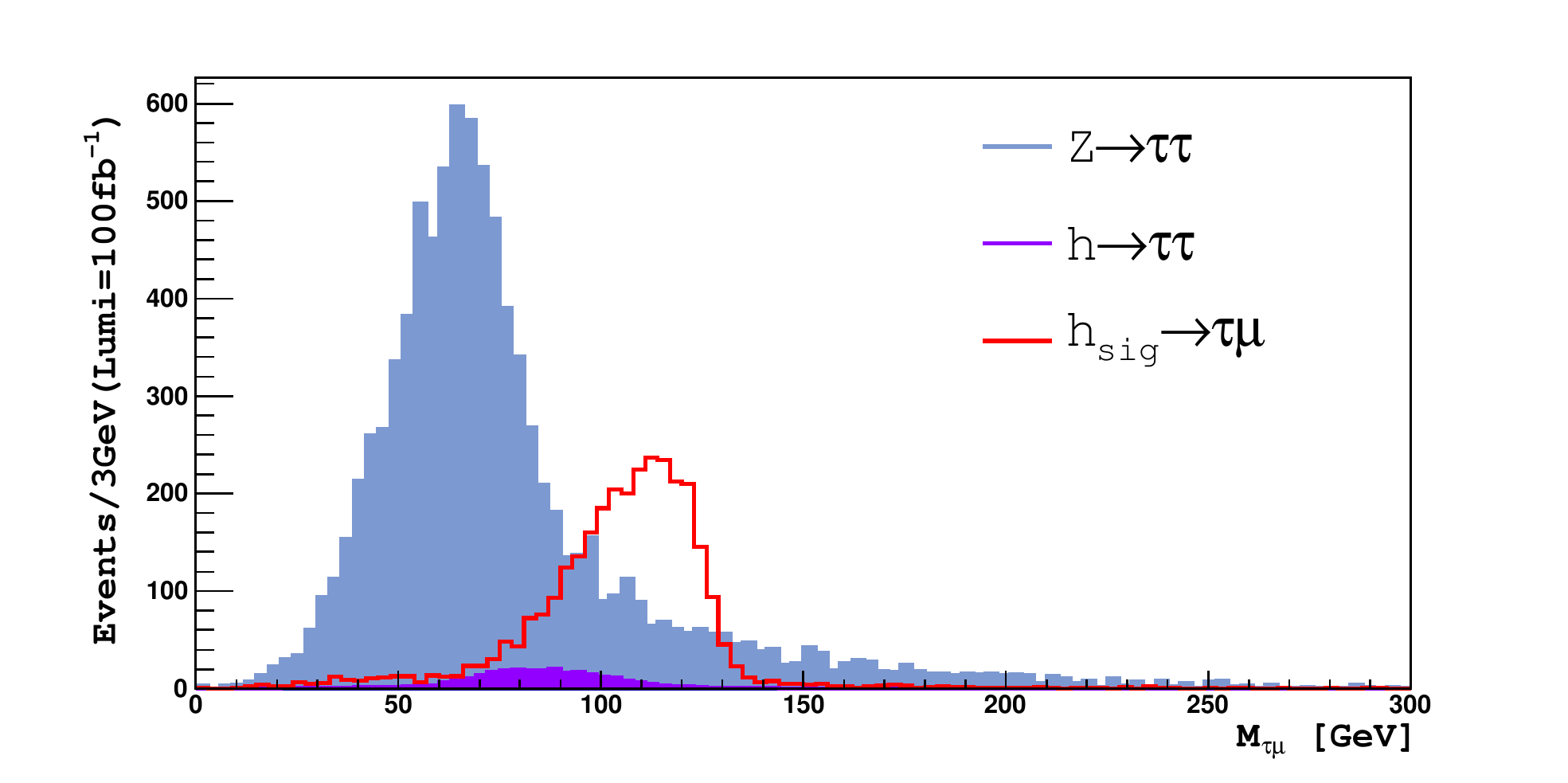}%
\includegraphics[width=8cm,height=5cm]{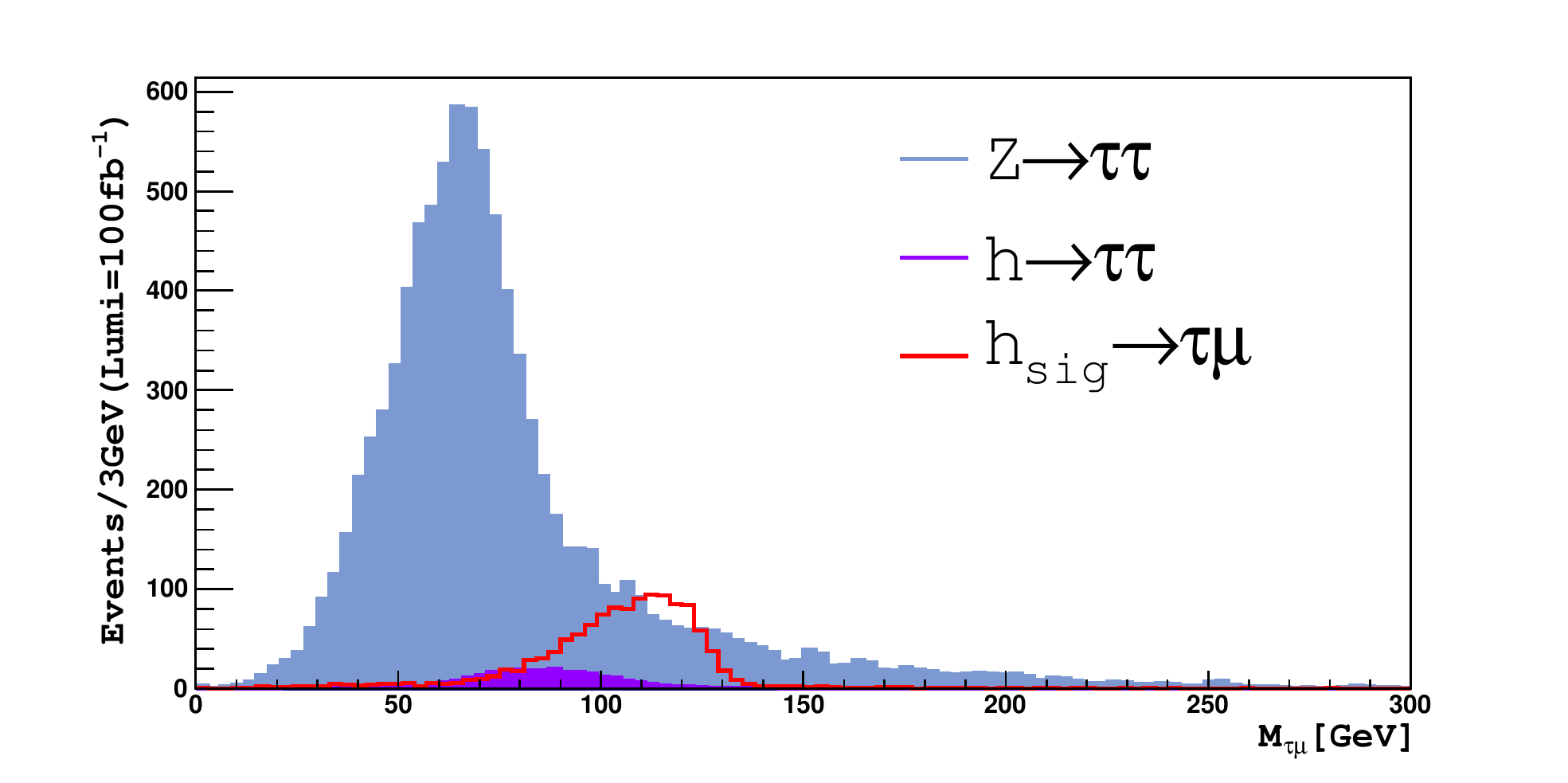}
\caption{Number of signal events for LFV $h\rightarrow \tau \mu$ decay versus the $\tau - \mu$ invariant mass ($M_{\tau\mu}$) at $\sqrt{s}=13$ TeV after $100$ fb$^{-1}$ luminosity for the solutions with the largest $BR(h\rightarrow \tau \mu) \sim 0.77\%$ (left), and light chargino solution ($m_{\tilde{\chi}_{1}^{\pm}}\sim 800$) with $BR(h\rightarrow \tau \mu) \sim 0.26\%$ (right). Also two dominant background processes are represented.}
\label{sigback}
\end{figure}

After we impose a veto for the processes with jets and/or missing $E_{T}$ in the final state, the main SM background is formed by the $Z\rightarrow \tau \tau$ and $h\rightarrow \tau \tau$ processes for the LFV Higgs decays into $\tau$ and $\mu$ leptons. We use MadGraph \cite{Alwall:2011uj} for matrix element calculation and event generation, in which the hadronisation effects are carried out by Pythia \cite{Sjostrand:2007gs}. In addition, the Pythia output is transferred to Delphes \cite{deFavereau:2013fsa} for the detector simulations. Finally, MadAnalysis5 \cite{Conte:2012fm} is employed for the data analyses.

Figure \ref{sigback} shows the number of signal events for LFV $h\rightarrow \tau \mu$ decay versus the $\tau - \mu$ invariant mass ($M_{\tau\mu}$) at $\sqrt{s}=13$ TeV after $100$ fb$^{-1}$ luminosity for the solutions with the largest $BR(h\rightarrow \tau \mu) \sim 0.77\%$ (left), and light chargino solution ($m_{\tilde{\chi}_{1}^{\pm}}\sim 800$) with $BR(h\rightarrow \tau \mu) \sim 0.26\%$ (right). The benchmark points are obtained from the scan over the BLSSM-IS parameter space represented in the previous section. The main contribution to the SM background comes from $Z\rightarrow \tau \tau$ whose peak is placed at about $M_{\tau\mu}\sim 75$ GeV. The contribution from $h\rightarrow \tau \tau$ is rather negligible and suppressed by the Z decay.  The left panel shows that the signal is likely being significantly above the background with the number of events about 250 for the solution with the largest $BR(h\rightarrow \tau \mu)$. The peak for this signal is placed at about $ M_{\tau\mu} \approx m_{h} \simeq 125$ GeV, which is well separated from the that of $Z\rightarrow \tau \tau$. On the other hand, the light chargino solution provides less number of events since it yields a lower $BR(h\rightarrow \tau \mu)$. As seen from the right panel, the number of signal ($\sim 100$) for this solution is slightly above the background. 

\begin{table}[h!]
\centering
\scalebox{0.9}{
\begin{tabular}{|c| c| c| c| c|}
\hline
\hline 
\#\# &parton level cuts &$P_T(l)\ge 30\&\eta\le 2.1$& $P_T(j)\ge 35\&\eta\ge 3.5$&Reject $(j\&MET)$ \\[0.5ex]
\hline
$Z\to\tau\bar{\tau}\ (bkg)$&408913&121249&16055&16055\\
$Wjj,W\to l\nu\ (bkg)$&24582&18160&11745&0.0\\
$Zjj,Z\to\tau\bar{\tau}\ (bkg)$&72694&41838&33678&12\\
$h\to\tau\bar{\tau}\ (bkg)$&3986&2574&848&848\\
$h\to\tau\bar{\mu}\ (sig)$&18062 (7225)&17165 (6866)&5257 (2103)&5257 (2103)\\
\hline
$S/\sqrt{S+B}$&24.9 (10.0)&38.3 (15.7)&20.2 (8.3)&35.3 (15.6)\\[1ex]
\hline 
\end{tabular}}
\caption{$h\to\tau\bar{\mu}$ Significance}
\label{table:significance}
\end{table}

Finally, we listed the number of events for the signals, and the significance over the relevant background processes. The number of events is the integration under the curves in plots in Figure \ref{sigback}. The cuts applied in our analyses are obtained first by ATLAS \cite{Aad:2015gha}, and the signal strength is defined as $S/\sqrt{S+B}$. The number of events for the first benchmark point with the largest $BR(h\rightarrow \tau \mu)$, and also the second benchmark point with relatively lighter chargino (in the paranthesis) are listed. The last line in Table \ref{table:significance} is the significance for these benchmark points. Surely, one can expect  the signals to be more significant with the increasing luminosity in the run of LHC.

\section{Conclusion}
\label{sec:conc}

We studied the LFV Higgs boson decay $h\to \tau \mu$ in several SUSY models such as MSSM, SSM, and BLSSM-IS. Even though this LFV Higgs boson decay can be induced through loop effects in the generic MSSM, in which the SSB slepton masses and/or trilinear scalar coupling can be non-universal such that the slepton families can mix each other, we found that $BR(h\rightarrow \tau \mu)$ can be realized at the order of $10^{-5}$ at most. Nevertheless, such solutions violate the constraints from another LFV process, $\tau \rightarrow \mu \gamma$, and if one imposes the constraints from these processes, the region with $BR(h\rightarrow \tau \mu) \gtrsim 10^{-8}$ is excluded by $\tau\rightarrow \mu \gamma$. Besides MSSM, we considered SSM in which the family mixing in the SSB slepton mass matrix and trilinear scalar interactions can be induced radiatively through the RGEs, even if one imposes the universal boundary conditions at $M_{{\rm GUT}}$. The LFV Higgs boson decay is mostly enhanced by the off-diagonal components of the slepton mass matrices, since the mixing in the trilinear coupling is negligible. In the SSM framework, we found that $\mu \rightarrow e \gamma$ brings the severest constraint on the parameter space, and $BR(h\rightarrow \tau \mu)$ is found at order of $10^{-10}$ at most without violating the constraints from the LFV processes, which is much smaller than the recent results reported by CMS and ATLAS.

On the other hand, our analyses showed that $h\rightarrow \tau \mu$ can be probed in the BLSSM-IS framework. $BR(h\rightarrow \tau \mu)$ can be as large as $0.77\%$ without violating the LFV constraints including $\mu \rightarrow e \gamma$. This is because the flavour mixing occurs mostly among the sparticles with opposite chirality (LR), while it is realized among the sparticles with the same chirality (LL or RR) in the generic MSSM and SSM. The form factors relevant to $h\rightarrow \tau \mu$ are enhanced with $\mu -$term, while large $\mu$ suppresses the possible contributions to $\mu\rightarrow e \gamma$ from the LR sector. Our results show that $BR(h\rightarrow \tau \mu)$ decreases in the region where $BR(\mu\rightarrow e \gamma)$ is enhanced. Besides, a large $BR(h\rightarrow \tau \mu)$ requires mostly heavy smuons and charginos which also lead to loop suppression in $\mu\rightarrow e \gamma$. On the other hand, in $h\rightarrow \tau \mu$, the chirality flip also takes a part and it is effective in compensating the heavy masses of sparticles running in loops. While smuon and chargino can be as heavy as multi-TeV, $m_{\tilde{\tau}} \lesssim 3$ TeV, $m_{\tilde{\chi}_{1}^{0}} \lesssim 0.8$ TeV, and $m_{\tilde{\nu}_{1}}\lesssim 2.5$ TeV are required to avoid the suppression from the heavy sparticles in $h\rightarrow \tau \mu$. In this context, BLSSM-IS type models can be accounted for a confirmed signal for the LFV Higgs boson decay. We also presented a signal simulation for a possible signal of $h\rightarrow \tau \mu$ over the relevant background processes with two benchmark points, and showed that the signal will be clearly detectable and its significance increases as higher luminosity is being collected during the LHC runs.

\section*{Acknowledgments}
The work is partially supported by the STDF project 13858,  the grant H2020-MSCA-RISE-2014 no. 645722 (NonMinimalHiggs), and 
 the European Union's Horizon 2020 research and innovation programme under the Marie Curie grant agreement No 690575. 
 A.H. is partially supported by the EENP2 FP7-PEOPLE-2012-IRSES grant.


\begin{thebibliography}{99}

\bibitem{Khachatryan:2015kon}
  V.~Khachatryan {\it et al.} [CMS Collaboration],
  Phys.\ Lett.\ B {\bf 749}, 337 (2015)
  doi:10.1016/j.physletb.2015.07.053
  [arXiv:1502.07400 [hep-ex]].


\bibitem{Aad:2015gha}
  G.~Aad {\it et al.} [ATLAS Collaboration],
  JHEP {\bf 1511}, 211 (2015)
  doi:10.1007/JHEP11(2015)211
  [arXiv:1508.03372 [hep-ex]];
  G.~Aad {\it et al.} [ATLAS Collaboration],
  JHEP {\bf 1511}, 211 (2015)
  doi:10.1007/JHEP11(2015)211
  [arXiv:1508.03372 [hep-ex]].

\bibitem{Arganda:2004bz} 
  E.~Arganda, A.~M.~Curiel, M.~J.~Herrero and D.~Temes,
  Phys.\ Rev.\ D {\bf 71}, 035011 (2005)
  doi:10.1103/PhysRevD.71.035011
  [hep-ph/0407302]; and references therein.

\bibitem{Adam:2013mnn} 
  J.~Adam {\it et al.} [MEG Collaboration],
  Phys.\ Rev.\ Lett.\  {\bf 110}, 201801 (2013)
  doi:10.1103/PhysRevLett.110.201801
  [arXiv:1303.0754 [hep-ex]].

\bibitem{Aubert:2009ag} 
  B.~Aubert {\it et al.} [BaBar Collaboration],
  Phys.\ Rev.\ Lett.\  {\bf 104}, 021802 (2010)
  doi:10.1103/PhysRevLett.104.021802
  [arXiv:0908.2381 [hep-ex]].

\bibitem{Calibbi:2012gr} 
  L.~Calibbi, D.~Chowdhury, A.~Masiero, K.~M.~Patel and S.~K.~Vempati,
  JHEP {\bf 1211}, 040 (2012)
  doi:10.1007/JHEP11(2012)040
  [arXiv:1207.7227].

\bibitem{Masiero:2004hg} 
  A.~Masiero, S.~K.~Vempati and O.~Vives,
  Nucl.\ Phys.\ Proc.\ Suppl.\  {\bf 137}, 156 (2004)
  doi:10.1016/j.nuclphysbps.2004.10.058
  [hep-ph/0405017].

\bibitem{Masiero:2002jn} 
  A.~Masiero, S.~K.~Vempati and O.~Vives,
  Nucl.\ Phys.\ B {\bf 649}, 189 (2003)
  doi:10.1016/S0550-3213(02)01031-3
  [hep-ph/0209303].

\bibitem{Paradisi:2006jp} 
  P.~Paradisi,
  JHEP {\bf 0608}, 047 (2006)
  doi:10.1088/1126-6708/2006/08/047
  [hep-ph/0601100].

\bibitem{Aloni:2015wvn} 
  D.~Aloni, Y.~Nir and E.~Stamou,
  JHEP {\bf 1604}, 162 (2016)
  doi:10.1007/JHEP04(2016)162
  [arXiv:1511.00979 [hep-ph]], and references therein.

\bibitem{Rosiek:1995kg} 
  J.~Rosiek,
  hep-ph/9511250.



\bibitem{Hollik:1988ii} 
  W.~F.~L.~Hollik,
  Fortsch.\ Phys.\  {\bf 38}, 165 (1990).
  doi:10.1002/prop.2190380302

\bibitem{Porod:2011nf}
  W.~Porod and F.~Staub,
  Comput.\ Phys.\ Commun.\  {\bf 183}, 2458 (2012).

\bibitem{Staub:2013tta} 
  F.~Staub,
  Comput.\ Phys.\ Commun.\  {\bf 185}, 1773 (2014)
  doi:10.1016/j.cpc.2014.02.018
  [arXiv:1309.7223 [hep-ph]].


\bibitem{Casas:2001sr} 
  J.~A.~Casas and A.~Ibarra,
  Nucl.\ Phys.\ B {\bf 618}, 171 (2001)
  doi:10.1016/S0550-3213(01)00475-8
  [hep-ph/0103065].


\bibitem{Khalil:2010iu} 
  S.~Khalil,
  Phys.\ Rev.\ D {\bf 82}, 077702 (2010)
  doi:10.1103/PhysRevD.82.077702
  [arXiv:1004.0013 [hep-ph]].

\bibitem{Elsayed:2011de} 
  A.~Elsayed, S.~Khalil and S.~Moretti,
  Phys.\ Lett.\ B {\bf 715}, 208 (2012)
  doi:10.1016/j.physletb.2012.07.066
  [arXiv:1106.2130 [hep-ph]].

\bibitem{Chatrchyan:2014nva} 
  S.~Chatrchyan {\it et al.} [CMS Collaboration],
  JHEP {\bf 1405}, 104 (2014)
  doi:10.1007/JHEP05(2014)104
  [arXiv:1401.5041 [hep-ex]].
  
\bibitem{Celis:2013xja} 
  A.~Celis, V.~Cirigliano and E.~Passemar,
  Phys.\ Rev.\ D {\bf 89}, 013008 (2014)
  doi:10.1103/PhysRevD.89.013008
  [arXiv:1309.3564 [hep-ph]].
  
\bibitem{Alwall:2011uj}
  J.~Alwall, M.~Herquet, F.~Maltoni, O.~Mattelaer and T.~Stelzer,
  JHEP {\bf 1106}, 128 (2011).  

\bibitem{Sjostrand:2007gs} 
  T.~Sjostrand, S.~Mrenna and P.~Z.~Skands,
  Comput.\ Phys.\ Commun.\  {\bf 178}, 852 (2008)
  doi:10.1016/j.cpc.2008.01.036
  [arXiv:0710.3820 [hep-ph]].

\bibitem{deFavereau:2013fsa}
  J.~de Favereau {\it et al.} [DELPHES 3 Collaboration],
  JHEP {\bf 1402}, 057 (2014).  
  
\bibitem{Conte:2012fm}
  E.~Conte, B.~Fuks and G.~Serret,
  Comput.\ Phys.\ Commun.\  {\bf 184}, 222 (2013).
  
\end{thebibliography}
\end{document}